\documentclass[aps,amsmath,amssymb,amsfonts,prd,twocolumn,groupedaddress,nofootinbib,floatfix,superscriptaddress]{revtex4-1}

\usepackage{graphicx,bm,color}
\usepackage{aas_macro}
\usepackage{multirow}
\usepackage{verbatim}
\usepackage{ulem}
\usepackage{mathtools}
\usepackage{colortbl}
\usepackage[dvipsnames]{xcolor}

\usepackage[pdftex,colorlinks,pdfpagelabels]{hyperref}
\hypersetup{
   bookmarksnumbered,
   citecolor={blue},
   linkcolor={blue},
   urlcolor={blue},
   filecolor={blue}
}

\begin{document}

\title[Galactic cosmic rays after the AMS-02 observations]{Galactic cosmic rays after the AMS-02 observations}

\author{Carmelo Evoli}
\email[]{carmelo.evoli@gssi.it}
\homepage[]{https://orcid.org/0000-0002-6023-5253}

\author{Roberto Aloisio}
\author{Pasquale Blasi}
\affiliation{Gran Sasso Science Institute, Viale F.~Crispi 7, 67100 L'Aquila, Italy}
\affiliation{INFN/Laboratori Nazionali del Gran Sasso, Via G.~Acitelli 22, Assergi (AQ), Italy}

\date{\today}

\begin{abstract}
The unprecedented quality of the data collected by the AMS-02 experiment onboard the International Space Station allowed us to address subtle questions concerning the origin and propagation of cosmic rays. Here we discuss the implications of these data for the injection spectrum of elements with different masses and for the diffusion coefficient probed by cosmic rays through their journey from the sources to the Earth. We find that the best fit to the spectra of primary and secondary nuclei requires (1) a break in the energy dependence of the diffusion coefficient at energies $\sim 300$ GV; (2) an injection spectrum that is the same for all nuclei heavier than helium, and different injections for both protons and helium. Moreover, if to force the injection spectrum of helium to be the same as for heavier nuclei, the fit to oxygen substantially worsens. Accounting for a small, $X_{s}\sim 0.4~\rm g~cm^{-2}$, grammage accumulated inside the sources leads to a somewhat better fit to the B/C ratio but makes the difference between He and other elements even more evident. The statistic and systematic error bars claimed by the AMS collaboration exceed the error that is expected from calculations once the uncertainties in the cross sections of production of secondary nuclei are taken into account. In order to make this point more quantitative, we present a novel parametrization of a large set of cross sections, relevant for cosmic ray physics, and we introduce the uncertainty in the branching ratios in a way that its effect can be easily grasped. 
\end{abstract}

\maketitle

\section{Introduction}
For decades the quest for better data has been constant in the field of cosmic ray (CR) physics. For the first time, at least in the energy region $E\lesssim$ TeV, the AMS-02 experiment onboard the International Space Station has reversed this situation: statistic and systematic errors on the measured spectra of protons, helium and other primary nuclei, as well as on secondary stable nuclei (boron, lithium, beryllium) are now at the few percent level, thereby providing an unprecedented framework for testing our ideas on the origin and transport of cosmic rays. 

On the other hand, our theoretical ability to make predictions on the spectra of nuclei, especially secondary nuclei, is limited by the uncertainties in the measured values of the cross sections, a point that has been raised by many authors \cite{Maurin2002,Strong2007,Tomassetti2017,Evoli2018}. 
The importance of this point can probably be best illustrated by using the case of boron: the boron-to-carbon ratio is routinely used to infer the mean grammage traversed by CRs while propagating in the Galaxy~\cite{Berezinskii1990}, but the reliability of the grammage depends on the knowledge of boron production cross sections from spallation of heavier elements and on the accuracy of the measurements of the fluxes of such elements (at roughly the same energy per nucleon).
Unfortunately the measured cross sections are known with at least $\sim 30\%$ error (even more for some channels) and the fluxes of elements heavier than carbon, oxygen and nitrogen remain rather uncertain. 

Some major breakthroughs have been made possible by the high precision measurements of AMS-02, first and foremost the detection of breaks in the spectra of virtually all nuclei, most likely hinting at a change of regime in the transport of Galactic CRs at rigidity $\sim 300$ GV. 
The anomalous hardening of the spectra of secondary stable nuclei also confirms that most likely the spectral breaks are related to CR transport rather than to subtle aspects of the acceleration process~\cite{Genolini2017}. 
The rising positron ratio~\cite{Aguilar2013} and the quasiconstant $\bar p/p$ ratio \cite{Aguilar2016} clearly represent major achievements of this experimental enterprise, with potentially huge implications for our theories on the origin of CRs, to the point that some authors \cite{Cowsik2016,Lipari2017} have put forward radically new ideas on the transport of CRs. Testing such ideas is extremely important, but to do so the first step is to understand whether there are serious problems in interpreting data on spectra of primary and secondary nuclei within standard assumptions.  

One such assumption, motivated by the fact that most our models for acceleration and transport of CRs are based on a strict rigidity dependence of both processes, is that the source spectra of all nuclei (whatever the sources may be) have the same general shape, especially at energies away from the injection energy and the maximum rigidity~\cite{Serpico2015}. This leads to the prediction that the fluxes of nuclei observed at the Earth should be different only because of interactions suffered during transport. It has been claimed~\cite{Adriani2011,Aguilar2015} that this is not the case and that proton and He injection spectra are required to be different. Less clear is whether the injection spectra of helium and heavier nuclei are required to be the same. 

The AMS-02 collaboration also provided the results of the measurement of the ratio of carbon and oxygen nuclei, both predominantly primary nuclei. The C/O ratio is expected to depart from a flat behavior at low energies due to two phenomena: (1) the mass of O nuclei is slightly larger than that of C nuclei, so that the corresponding spallation cross section is somewhat larger, so as to make the O nuclei more depleted at energies where spallation is relevant; (2) about 20\% of the flux of carbon at low energies is due to spallation of O nuclei. These two phenomena are responsible for a C/O ratio that decreases with energy below $\sim 100$ GeV/n. This trend depends on the same grammage that is probed through observations of the B/C ratio, and can be considered as an important test of consistency. At the same time, it is important to stress that among the nuclei whose flux has been measured so far, only protons and oxygen can be considered as truly primary, in good approximation. As stressed above, even carbon is polluted by a sizeable secondary contribution from spallation of oxygen and other heavier elements. This observational situation is unprecedented and allows us to test for the first time the essential aspects of CR transport in the Galaxy and perhaps seek signs of possible failure of our basic ideas.   

The article is structured as follows: in \S~\ref{sec:propa} we describe the basic aspects of our propagation model, including relevant cross sections. The methodology adopted in this work is illustrated in \S~\ref{sec:method}. In \S~\ref{sec:results} we discuss our main results. Conclusions are provided in \S~\ref{sec:conclude}. A complete list of relevant measured cross sections and fits to the data are provided in Appendix~\ref{sec:appendix}.

\section{Propagation model}
\label{sec:propa}

If we restrict our attention to primary nuclei and stable secondary nuclei the problem of cosmic ray (CR) transport can be well described by using a 1D advection-diffusion equation including the whole chain of spallation reactions from heavier nuclei to lighter nuclei. {More complex treatments of diffusion are not necessarily more realistic: for instance three dimensional diffusion models are often used, but the different transport parallel and perpendicular to the local direction of the magnetic field is not accounted for. Moreover, the low level of measured CR anisotropy suggests that the radial dependence in the distribution of sources does not play a large role. Finally, the CR flux observed at the Earth typically comes from sources at distances comparable with the size of the halo, $H\sim 4$ kpc. On such scales the CR gradient inferred from gamma ray observations is small, again suggesting that a 1D modelling of the transport is sufficient for our purposes.}

We also assume for simplicity that the disc is much thinner than the magnetized halo. Problems may appear in this approach when the loss length of nuclei of type $\alpha$ is smaller than the thickness of the disc, which may occur for unstable nuclei such as $^{10}$Be at energies $\lesssim 1$ GeV/n. Again, we do not consider here such low energies, where other effects contribute to make observations of difficult interpretation. The general form of the transport equation describing this situation \cite{Jones2001,Aloisio2013}, known as the modified weighted slab model, reads as follows:
\begin{multline}\label{eq:slab}
-\frac{\partial}{\partial z} \left[D_{\alpha}(p) \frac{\partial f_{\alpha}}{\partial z}\right] 
+ u \frac{\partial f_{\alpha}}{\partial z}
- \frac{du}{dz} \frac{p}{3} \frac{\partial f_{\alpha}}{\partial p} 
\\
+ \frac{1}{p^{2}} \frac{\partial}{\partial p}\left[ p^{2} \left(\frac{dp}{dt}\right)_{\alpha,ion} f_{\alpha}\right] 
+ \frac{\mu v(p) \sigma_{\alpha}}{m}\delta(z) f_{\alpha} 
\\ 
= 2 h_d q_{0,\alpha}(p) \delta(z) 
+ \sum_{\alpha'>\alpha} \frac{\mu\, v(p) \sigma_{\alpha'\to\alpha}}{m}\delta(z) f_{\alpha'},
\end{multline}
where $f_\alpha (p,z)$ is the phase-space density of the CR species $\alpha$ as a function of the particle momentum $p$ and position $z$ away from the disc, $v(p)=\beta(p) c$ is the velocity of a nucleus, and $\mu$ is the surface density of the disc.

The terms in the left-hand side (LHS) of Eq.~\ref{eq:slab} describe particle diffusion, advection, energy losses and spallation, respectively. 

We assume that the diffusion coefficient is spatially constant and the only function of the particle momentum. Moreover, assuming that the observed spectral hardening at $\sim 300$ GV is due to a change of regime in particle diffusion \cite{Genolini2017}, we adopt the following functional form for the dependence of the diffusion coefficient on rigidity $R$:
\begin{equation}
D(R) = \beta D_0 \frac{(R/{\rm GV})^\delta}{[1+(R/R_b)^{\Delta \delta /s}]^s}, 
\end{equation}
where $D_0$ is the value of the diffusion coefficient at $R=1$ GV and the break is described in terms of the parameters $s$, $\Delta \delta$, $R_b$ which are, respectively, the smoothing, the magnitude and the characteristic rigidity of the break. While this is clearly a very phenomenological approach, this type of energy dependence has also been derived in more physics motivated scenarios: for instance the transition from self-generated turbulence to preexisting turbulence \cite{Blasi2012,Aloisio2013,Evoli2018b} and a nonseparable spatially dependent diffusion coefficient in the Galactic halo \cite{Tomassetti2012} both lead to a break in the {\it effective} diffusion coefficient. 

The second term on the LHS of Eq.~\ref{eq:slab} accounts for particle advection with velocity $u$, which may describe the presence of a Galactic wind if one is present or advection with Alfv\'en waves if it happens that there are more waves moving outward than moving inward (this would be the case if the waves are self-generated through streaming instability excited by CRs themselves~\cite{Blasi2012}). 
In the simplest scenario, the direction of the advection velocity is expected to reverse direction above and below the disc, so that $du/dz = 2u \delta(z)$, which determines the adiabatic losses through the third term on the LHS. 

{Since we get inspiration from models in which waves are self-generated, it is worth mentioning that such models are, in general, incompatible with having second order Fermi acceleration (reacceleration) in the ISM, which is in fact not included in Eq.~\ref{eq:slab}. Reacceleration requires the presence of waves moving in both directions, so that the mean Alfv\'en speed is low or vanishing (no advection with the waves), while self-generated waves all move away from the disc, in the direction of decreasing CR flux. In such models, for low enough energies (typically $\lesssim 10$ GeV) transport becomes advection dominated. It is also worth recalling that some recent work \cite{Drury2017} hints at serious energetic problems that reacceleration might run into, while being also hard to reconcile with constraints derived from radio observations \cite{Strong2011,DiBernardo2013}.}

Ionization energy losses are taken into account through the fourth term on the LHS of Eq.~\ref{eq:slab}, where 
\begin{equation}
\left(\frac{dp}{dt}\right)_{\alpha,ion} = 2 h_d \dot{p}_{0,\alpha} \delta(z),
\end{equation}
valid in the assumption that the disc is infinitely thin. The function $\dot{p}_{0,\alpha}$ is the same as used in Ref.~\cite{Mannheim1994}.  

The spallation of CR nuclei is treated as an effective {\it sink} term, $f_{\alpha}/\tau_{sp,\alpha}$, with a rate that is proportional to the spallation cross section $\sigma_{\alpha}$ and the gas density in the interstellar medium. More accurately, one can write the spallation rate taking into account the fact that the ISM target gas is mainly made of hydrogen (H) and helium (He): 
$$
\tau_{sp,\alpha}^{-1} = v(p) \left( n_{\rm H} \sigma_{\alpha}^{(\rm H)} + n_{\rm He} \sigma_{\alpha}^{(\rm He)}\right) = \frac{\mu v \sigma_{\alpha}}{m} \delta(z),
$$
where we introduced the mean mass $m=m_{p}\frac{1+4f_{\rm He}}{1+f_{\rm He}}$ and the effective spallation cross section 
$$
\sigma_{\alpha}=\sigma_{\alpha}^{(\rm H)}\frac{1+4^{2/3}f_{\rm He}}{1 + f_{\rm He}}.
$$

The quantity $f_{\rm He}=n_{\rm He}/n_{\rm H}=0.1$~\cite{Ferriere2001} is the fraction of helium in the ISM with respect to hydrogen. In the expression for the cross section, the factor $4^{2/3}$ reflects the assumption that the cross section for spallation on He is geometrically larger than that on hydrogen: $\sigma_{\alpha}^{(\rm He)}=4^{2/3}\sigma_{\alpha}^{(\rm H)}$. 
Finally we introduced the effective grammage of the disc (surface density) $\mu = 2h_{d} m n_{H} (1+f_{\rm He})$. From observations $\mu \simeq 2.3$~mg/cm$^2$~\cite{Ferriere2001}. 
A similar functional form describes the spallation of a nucleus of type $\alpha'>\alpha$ into a nucleus of type $\alpha$ (last term on the RHS of Eq.~\ref{eq:slab}).

For the {\it total} spallation cross sections on a hydrogen target we followed the results of~\cite{Tripathi1996,Tripathi1997} that allow a good fit to all existing data and differ from older phenomenological approaches by at most 5$\%$-10$\%$. 
The situation is more critical in terms of production cross sections of secondary isotopes (e.g., isotopes of Li, Be, B) from spallation of a heavier nucleus.
{Sporadic measurements of relevant reactions (although at energies much below GeV/n) were already available in the 1960s (see, eg.,~\cite{Bernas1965,Yiou1968}). 
The first systematic measurements of the most relevant secondary production channels were made from the 1970s till the end of the 1990s. 
Based on these results, an attempt was made to establish semiempirical~\cite{Silberberg1973,Silberberg1998} or fullyempirical parametrizations~\cite{Webber2003} to evaluate the cross sections for any given spallation channel and energy. 
These parametrizations, although with a questionable accuracy, are built by capturing some global trends (for instance scaling relations involving the number of neutrons or the difference in mass between the projectile and the fragment) directly inferred from the data. A major problem is that most of the experimental data of production channels are at low energies, usually at hundreds of MeV/n, and just for some channels are available at few GeV/n's.
Moreover, systematic uncertainties associated to these measurements (especially the older ones) are difficult to assess and the contribution of {\it ghost} nuclei (whose lifetime is long enough not to decay during the measurements but much shorter than the CR escape time) is practically unexplored~\cite{Maurin2002}. 
Building on these previous works, the GALPROP collaboration developed a comprehensive set of routines that combined various datasets of measurements with the parametrizations of previous works, eventually providing original fits to the data for some specific reactions~\cite{Strong2007}.

In our work we follow the formalism described in~\cite{Evoli2018} where the \emph{direct} spallation cross sections on hydrogen target are evaluated by normalizing the Webber~\cite{Webber2003} and the Silberger and Tsao~\cite{Silberberg1998} formulas to available data.
Unstable elements with a short lifetime compared with the typical propagation times scales are considered as instantaneous source terms to the daughter nucleus that they decay into. The Webber model has been used to compute that contribution to each channel, including about 150 reactions.

\begin{figure*}[t]
\begin{center}
\includegraphics[width=0.98\columnwidth]{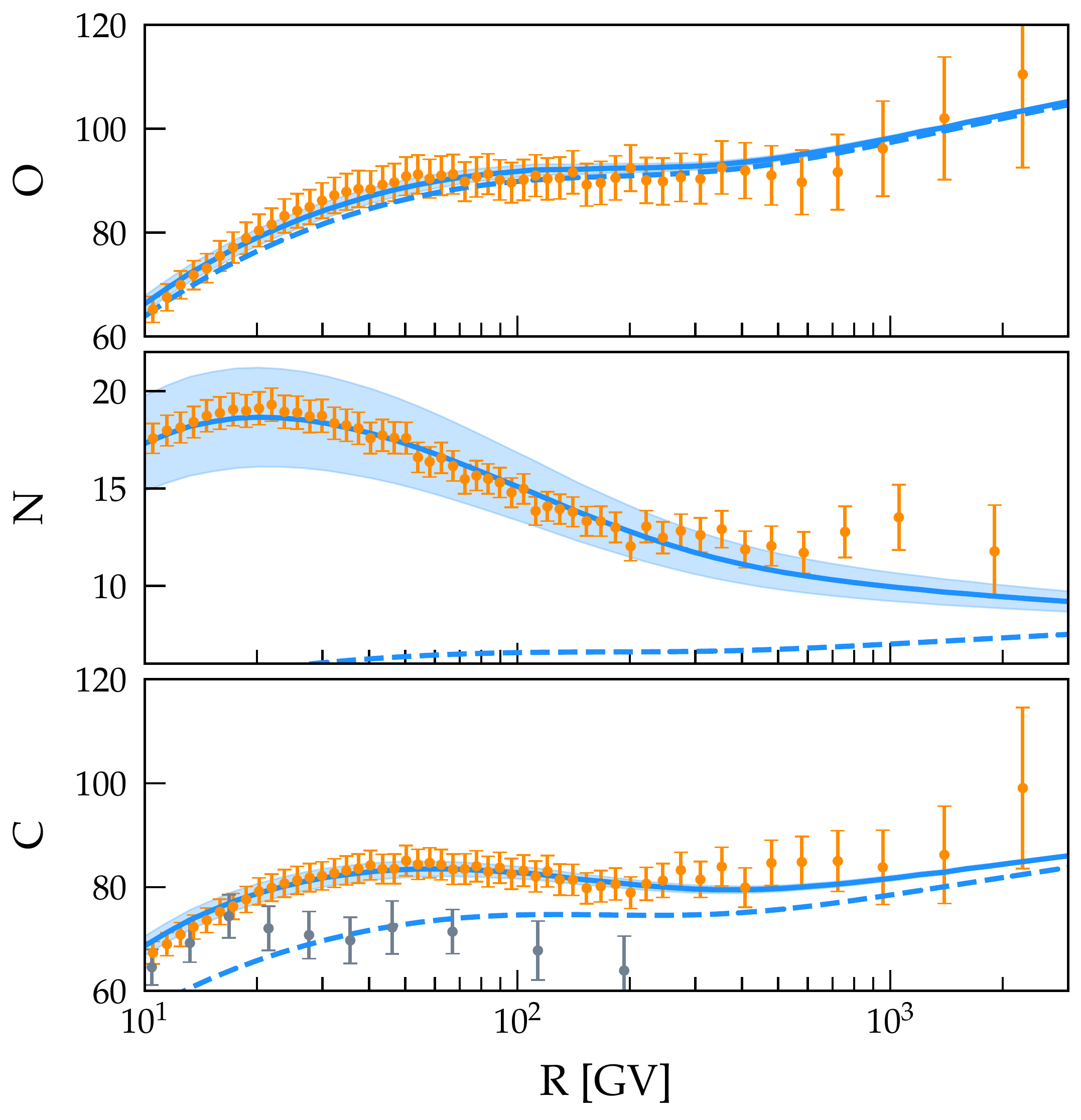}
\hspace{\stretch{1}}
\includegraphics[width=0.98\columnwidth]{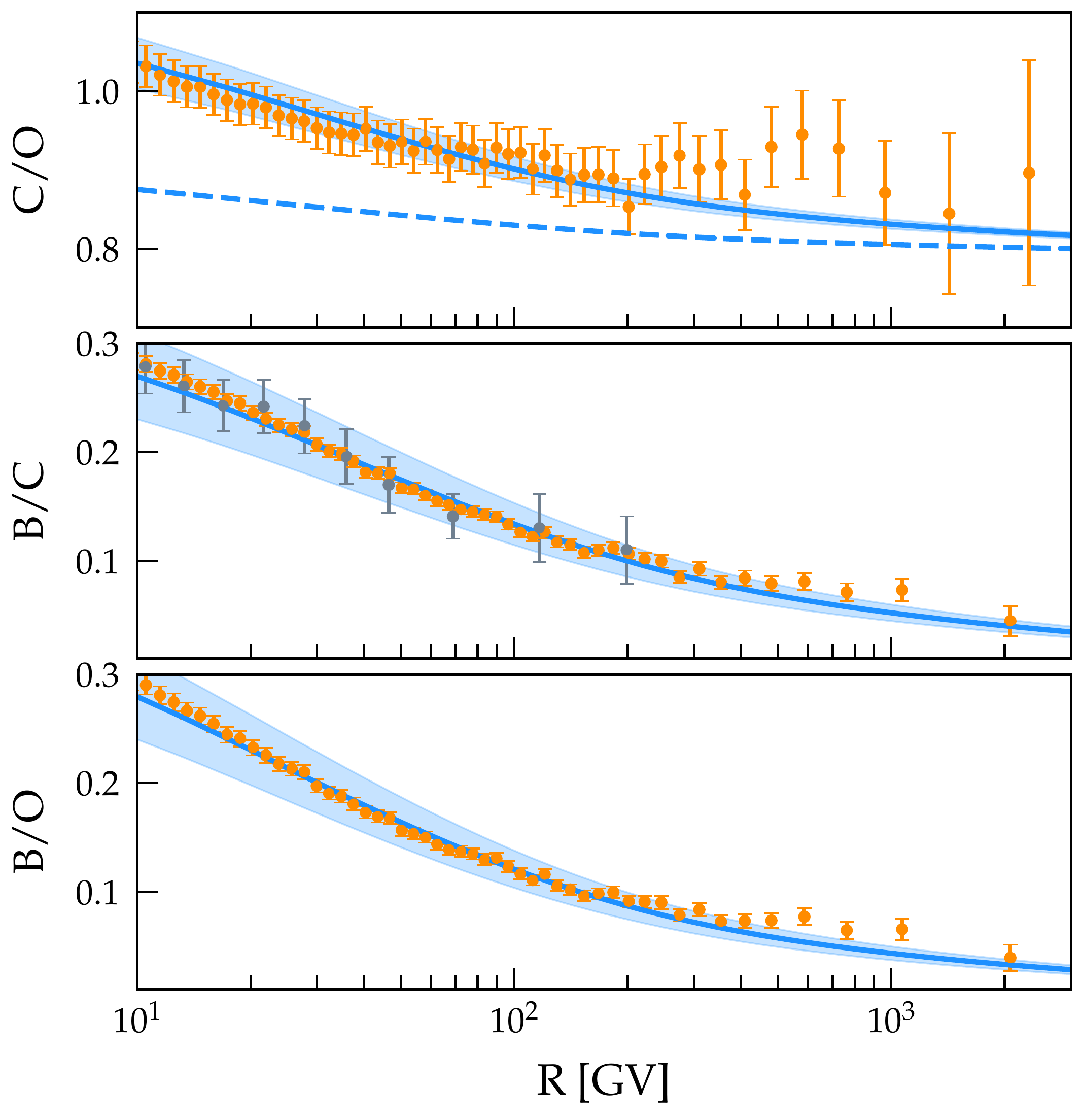}
\end{center}
\caption{The best-fit model as described in \S~\ref{sec:A} compared with AMS02 (orange) and PAMELA (gray) data. Shaded regions represent the 1-$\sigma$ uncertainty region associated to cross sections. The dashed lines show the contribution from primary sources $q_{\alpha}$ only. Left panel (from top to bottom): The oxygen, nitrogen and carbon fluxes (in units of GV$^{-1}$~m$^{-2}$ s$^{-1}$ sr$^{-1}$) as a function of rigidity $R$ and multiplied by $R^{2.7}$. Right panel (from top to bottom): The C/O, B/C and B/O ratios.}
\label{fig:bestfit}
\end{figure*}

In this paper, we update the evaluation of the most relevant channels for Li, Be and B production by performing new phenomenological fits to a larger set of data as discussed in the Appendix~\ref{sec:appendix}. 
Additionally, these fits allow us to assess the uncertainty on the determination of the high-energy values of these cross sections and therefore to better estimate their impact on the secondary-over-primary ratios as discussed in the next section.
We notice that such partial cross sections have larger uncertainties, that for some channels reach $\gtrsim 50\%$. 
Limited to the production of secondary $^{3}$He and $^2$H we make use of the results published in~\cite{Coste2012} by implementing in our calculations the cross sections distributed within the USINE code\footnote{\url{https://lpsc.in2p3.fr/usine}}.}

Finally, particles are injected in the disk at a rate $q_{0,\alpha}$ that we compute assuming that all CRs are accelerated in SNRs. The slope of the power law source spectrum is assumed to be the same for all sources, $\gamma$. The injection efficiency with respect to the total kinetic energy of the supernova explosion is $\epsilon_\alpha$. The source term can be written as:
\begin{equation}
q_{0,\alpha} = \frac{\epsilon_\alpha E_{\rm SN}{\mathcal R}_{\rm SN}}{\pi R_d^2 \Gamma(\gamma)c(m_pc)^4} \left( \frac{p}{m_p c} \right)^{-\gamma},
\end{equation}
where $E_{\rm SN} = 10^{51}$~erg is the total kinetic energy of a supernova, $R_d = 10$~kpc is the radius of the Galactic disk, ${\mathcal R}_{\rm SN} = 1/30$~yr$^{-1}$ is the rate of SN explosions and $\Gamma(\gamma) = 4\pi \int_0^\infty dx x^{2-\gamma} [\sqrt{x^2 + 1} - 1]$. 

Following the procedure outlined in Refs.~\cite{Jones2001,Aloisio2013}, after imposing the boundary condition that $f_\alpha(p, z=\pm H) = 0$, one can transform Eq.~\ref{eq:slab} in a modified weighted slab transport equation:
\begin{multline}\label{eq:slab2}
\frac{I_{\alpha}(E)}{X_{\alpha}(E)} + \frac{d}{dE}\left\{\left[ \left(\frac{dE}{dx}\right)_{\rm ad} +  \left(\frac{dE}{dx}\right)_{\rm ion, \alpha}\right] I_{\alpha}(E)\right\} \\ + \frac{\sigma_{\alpha} I_{\alpha}(E)}{m} = 2 h_d \frac{A_{\alpha} p^{2} q_{0,\alpha}(p)}{\mu\, v} + \sum_{\alpha'>\alpha} \frac{I_{\alpha}(E)}{m}\sigma_{\alpha'\to\alpha},
\end{multline}
where $I_{\alpha}(E) = A_{\alpha}p^{2} f_{0,\alpha}(p)$ is the flux of nuclei of type $\alpha$ with kinetic energy per nucleon $E$. 
We introduced the quantity
\begin{equation}
X_{\alpha}(E) = \frac{\mu \,v}{2 u} \left[ 1-\exp\left(-\frac{uH}{D_{\alpha}(E)}\right)\right],
\end{equation}
that represents the grammage for nuclei of type $\alpha$, while
\begin{equation}
\left(\frac{dE}{dx}\right)_{\rm ad} = -\frac{2 u}{3\mu\, c} \sqrt{E(E+2m_p c^2)} 
\end{equation}
and
\begin{equation}
\left(\frac{dE}{dx}\right)_{\rm ion, \alpha} = -\frac{2h_d}{\mu} \dot{p}_{0,\alpha}
\end{equation}
are the rate of adiabatic energy losses due to advection and ionization, respectively. 

Eq.~\ref{eq:slab2} is solved for each species numerically following the formal solution discussed in Ref.~\cite{Aloisio2013}.
Since the lighter species originate from the spallation of the heavier ones, we start the evaluation of CR densities from heavy nuclei and proceed toward lighter ones, using the spallation of heavier species as a source term for lighter nuclei. This procedure is repeated for all nuclei in reverse mass order, starting from iron ($Z=26$) and all the way down to hydrogen. The injection efficiency $\epsilon_\alpha$ for each type of primary nucleus is tuned to fit observations. For nuclei heavier than oxygen where AMS-02 data are not yet available, we fit the normalization to CREAM data~\cite{Ahn2009}. 

\section{Methodology}
\label{sec:method}

If the spectral breaks observed for protons and helium nuclei originate due to a change in the properties of diffusion in Galactic magnetic fields, as we assume throughout this article, the change of slope in the diffusion coefficient $\Delta\delta$, the smoothness $s$ of the transition between the two regimes and the energy/rigidity ($R_{b}$) where the transition occurs can all be derived from analyzing the proton spectrum alone, since it is little affected by phenomena other than diffusion, at least for energies $\gtrsim 10$ GeV. These quantities can be fitted independent of the injection spectrum and the local slope of the diffusion coefficient. We use the AMS-02 data for protons~\cite{Aguilar2015b} and derived that $s = 0.1$, $\Delta \delta = 0.2$ and $R_b = 312$~GV. Throughout the rest of this article we will adopt these reference values, and we will assume that they also apply to the transport of all nuclei other than hydrogen, based on the fact that diffusion only depends upon particle rigidity. 
On the other hand, the normalization of the diffusion coefficient $D_{0}$ (namely the grammage) and its absolute energy dependence can only be derived from indicators including production of secondary nuclei, such as B/C and B/O, or even C/O since a sizeable fraction of carbon nuclei in the cosmic radiation has a secondary origin due to spallation of oxygen nuclei. Below, the symbol $\delta$ will be used to denote the slope of the diffusion coefficient for rigidity below $R_{b}$, while the slope above the break is always $\delta-\Delta\delta$.  Since the fluxes of stable nuclei only depend upon the ratio $D_0/H$, for the purpose of numerical calculations we adopt $H=4$ kpc. 

The CR fluxes at low rigidity ($\lesssim 50$ GV) are affected by solar modulation and by advection. We treat solar modulation in the force field approximation with a Fisk potential $\phi$ that is one of the parameters we find from the fit to data. In order to minimize the impact of unknowns in solar modulation (for instance data for different nuclei may be collected in different periods and therefore reflect different values of $\phi$), we decided to marginalize the results of the fit to the data in the region $R>20$ GV. Moreover the boron flux that we heavily use in the following, for rigidity $\lesssim 20$ GV is affected by the decay of $^{10}$Be that is not included in the following treatment because of complications deriving from the fact that the decay occurs inside the Galactic disc and the weighted slab model is inappropriate to describe such situation.  

\section{Results}
\label{sec:results}

The calculation described above is used here to determine the transport properties (diffusion and advection) that best describe data and to assess the existence of possible deviations from the standard scenario of CR transport in the Galaxy. The fits are carried out by minimizing the $\chi^{2}$ over a parameter space made of $D_{0}$, $\delta$, the slope $\gamma$ and the normalizations $\epsilon_\alpha$ of the injection spectra of nuclei, the advection velocity $u$ and the Fisk potential $\phi$. The minimization procedure is based on MINUIT~\cite{James1975}. 

Below we will illustrate three parts of the calculation: 1) we first minimize the $\chi^{2}$ of the fit to the spectra of C, N and O and B/C and determine the transport properties. In this case we show the inferred spectra of H and He obtained assuming that the source spectrum is the same as that of nuclei. 2) We repeat the same calculation but including the He spectrum in the fit and requiring that He is injected with the same spectrum as heavier nuclei, which is what one would expect in the case of pure rigidity dependent acceleration. 3) We assess the role of a possible, energy independent grammage accumulated by CRs inside the sources, during the acceleration process. 

\subsection{Fit to nuclei heavier than He}
\label{sec:A}

The spallation process is more effective for heavier nuclei at a given energy per nucleon, being the cross section approximately $\propto A^{2/3}$. As a result, oxygen nuclei are destroyed slightly faster than carbon nuclei. Moreover the spallation of $^{16}$O partially results in the production of $^{12}$C. These two factors lead to a C/O ratio that is expected to decrease with energy for $R\gtrsim 100$ GV, despite the fact that C and O are typically considered as primary nuclei. The B/C ratio also decreases with energy as a result of the B production in spallation events involving mainly C and O nuclei. Hence the two ratios, B/C and C/O are both expected to contain information about the grammage traversed by CRs during transport through the Galaxy. 
Here we calculate the best combined fit to the fluxes of C, N and O and the B/C and C/O ratios, so as to infer the source spectrum of nuclei and their transport properties. 

The error bar on each data point has been calculated by summing in quadrature the statistical and systematic uncertainties, as quoted by the AMS-02 collaboration. 
We also assume that systematic errors are completely uncorrelated. 

\begin{figure}[t]
\begin{center}
\includegraphics[width=0.98\columnwidth]{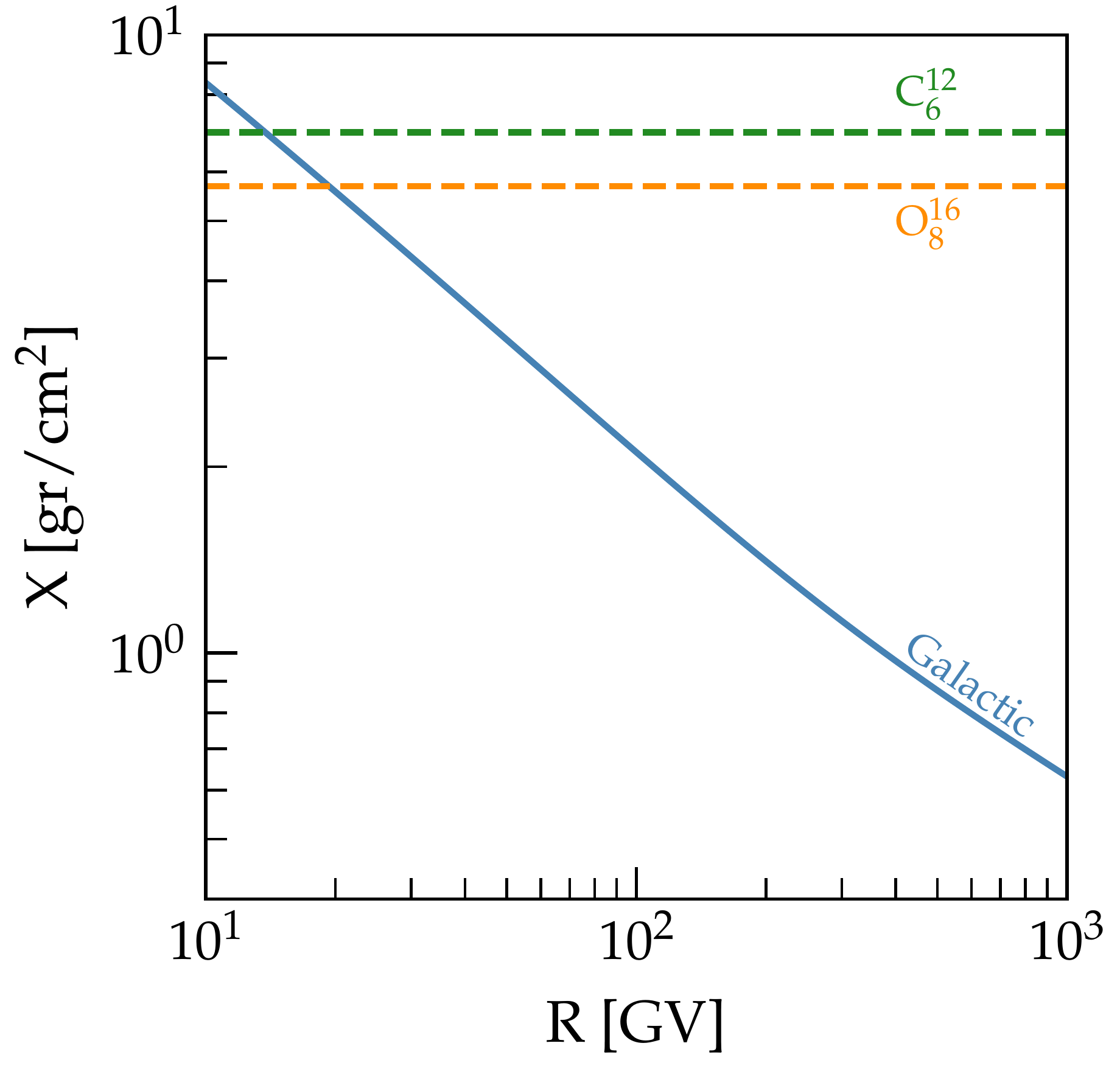}
\end{center}
\caption{The grammage corresponding to the best fit model described in \S~\ref{sec:A} as a function of rigidity is shown as a solid blue line. The dashed lines refers to carbon (green) and oxygen (orange) inelastic critical grammage.}
\label{fig:grammage}
\end{figure}

\begin{figure}[t]
\begin{center}
\includegraphics[width=0.98\columnwidth]{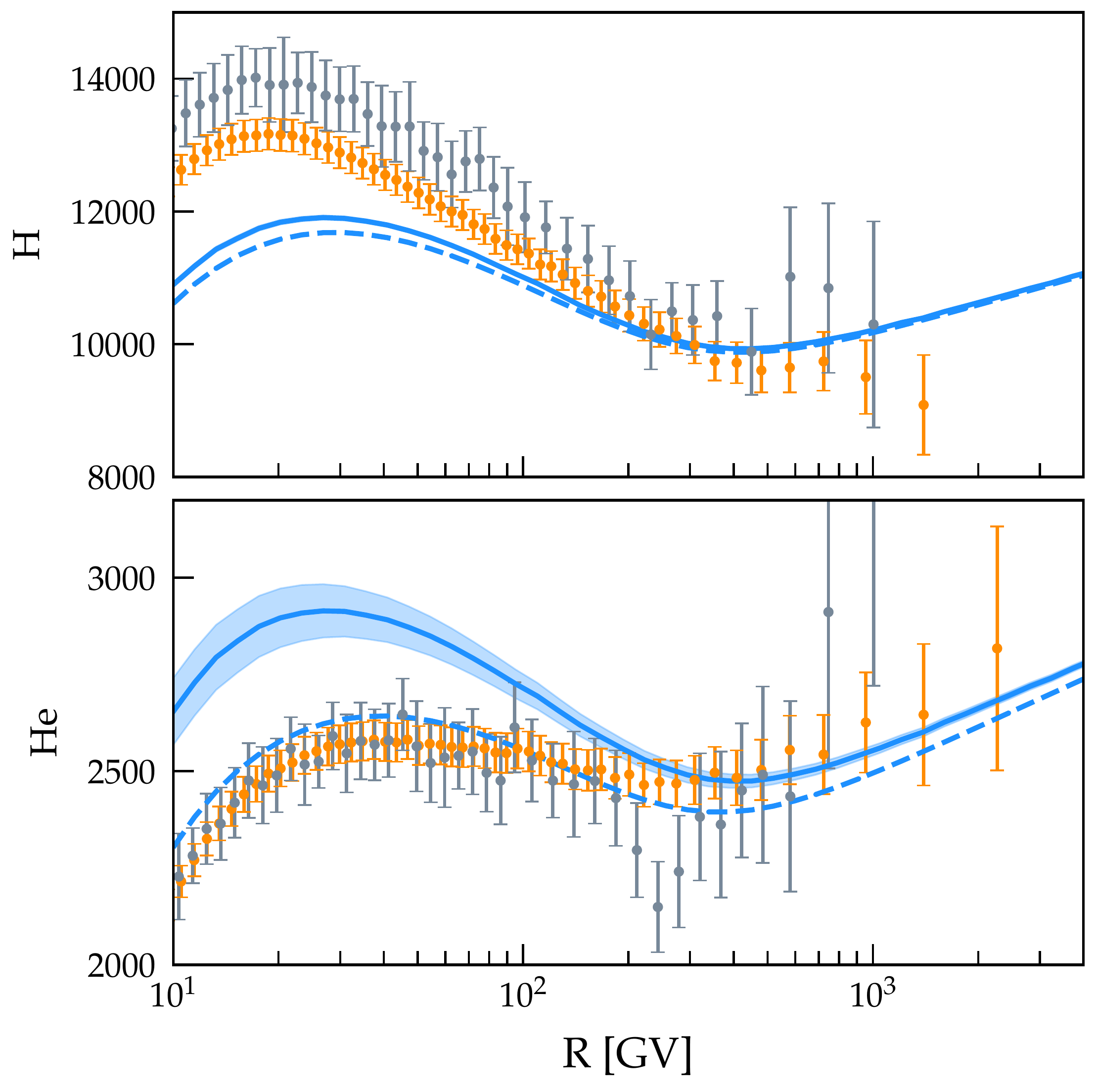}
\end{center}
\caption{Proton and helium spectra as measured by AMS02 (orange) and PAMELA (gray). The solid lines are the prediction of our model in which the injection slope and the transport parameters are inferred by heavier than He species (\S~\ref{sec:A}). Dashed lines show the primary contribution of $^1$H and $^4$He. Flux units as in Fig.~\ref{fig:bestfit}.}
\label{fig:bestfit_HHe}
\end{figure}

Our best fit scenario is illustrated in Fig.~\ref{fig:bestfit} and corresponds to the following values of the parameters: $u = 7$~km/s, $D_0 = 1.1 \times 10^{28}$~cm$^2$/s, $\delta = 0.63$ and $\gamma = 4.26$ (the injection spectrum of all nuclei is assumed to be the same). The modulation potential is $\phi = 0.51$ GV. The left panel of Fig.~\ref{fig:bestfit} shows our results for the spectra of C, N and O, while the right panel shows the C/O, B/C and B/O ratios. The data points are from AMS-02 and PAMELA (grey data points), when available. The dashed lines represent the results of our calculations if the contribution to the given quantity from spallation of heavier elements were neglected. 

A few findings emerge from the left panel of Fig.~\ref{fig:bestfit}: Oxygen is the nucleus that best satisfies the definition of primary nucleus, since the contribution to $O$ from spallation of heavier elements is smaller than the error bars in the data. For C nuclei this is clearly not true: at $R\lesssim 200$ GV the secondary contribution to C is $\sim 20\%$ (mainly from fragmentation of O nuclei). For this reason, one should not expect that the scaling B/C$\sim X(R)$ holds very well, which is in fact the case. In this sense, the B/O ratio is probably a better indicator of the grammage $X(R)$ traversed by CRs.

This consideration is even more true for nitrogen, which is roughly half primary and half secondary product of spallation of heavier elements. 
The bulk of the N flux below $\sim 100$ GV is in fact of secondary origin. The good fit to C, N and O fluxes makes us confident that the general picture of transport of these nuclei is in fact self-consistent, with no particularly evident anomaly arising from the comparison between data and calculations. 

Given the high accuracy of the AMS-02 data, our analysis can assess the role played by the uncertainties in the spallation cross sections for the predicted fluxes. In order to do so, we have repeated our calculations one hundred times for each set of parameters, extracting each time the spallation cross sections from a Gaussian distribution with central value at the best fit and a width which is assumed to be $\sim 5\%$ for the total cross sections and $\sim 30\%$ for the partial cross sections. The results are shown as shaded areas in Fig.~\ref{fig:bestfit}. This exercise shows how the uncertainties in the cross sections affect the results of transport calculations and how such uncertainties compare with the AMS-02 error bars. 
While not dramatic for the fluxes of C and O, the implications for the nitrogen flux are clear, especially at low energies. 

The effects of the uncertainties in the partial cross sections on the secondary nuclei are evident from the right panel of Fig.~\ref{fig:bestfit}. The uncertainties in the B/C and the B/O ratios as due to cross sections are much larger than the error bars in AMS-02 data, so that further accuracy in measuring such quantities would not help in better constraining the characteristics of CR transport. The top panel on the right side of Fig.~\ref{fig:bestfit} shows the C/O ratio, which is very well described as due to spallation of O nuclei into C nuclei, provided the two are injected with the same spectrum. Accounting only for O spallation but neglecting its contribution to the flux of C would result in the dashed line, namely it would fail to describe the C/O ratio. The uncertainties in the partial cross section O$\to$C does not hinder such a conclusion, as one can see from the limited size of the shaded region. 

The grammage corresponding to the best fit discussed with reference to Fig.~\ref{fig:bestfit} is shown in Fig. \ref{fig:grammage}, where we also plot the critical grammage, $X_c = \sigma_\alpha / m$, for $^{12}$C and $^{16}$O spallation (horizontal lines). 

The picture that emerges from this first part of our calculation is rather encouraging in that it shows that a self-consistent description of the nuclei, all injected with the same spectrum, is possible and in fact the only differences that arise from observations of the spectra of nuclei can be accounted for in terms of the different levels of nuclear fragmentation that they suffer. The only point that is left to check is whether the same parameters can be used to describe the observed spectrum of He and, if possible, even protons. The spectra of H (including deuterium) and He (sum of $^{4}$He and $^{3}$He) are shown in Fig.~\ref{fig:bestfit_HHe} in the top and bottom panel respectively. For protons we also account for the contribution of secondary deuterium coming from spallation of heavier elements and secondary protons produced in inelastic interactions of primary protons ($p+p\to p+p+\rm Anything$). The spectrum of protons without these additional contributions is shown as a dashed line in the top panel. There is no doubt that the observed proton spectrum cannot be properly described unless the injection spectrum of protons is assumed to be softer than that of nuclei by $\sim 0.05$.  

\begin{figure*}[t]
\begin{center}
\includegraphics[width=0.98\columnwidth]{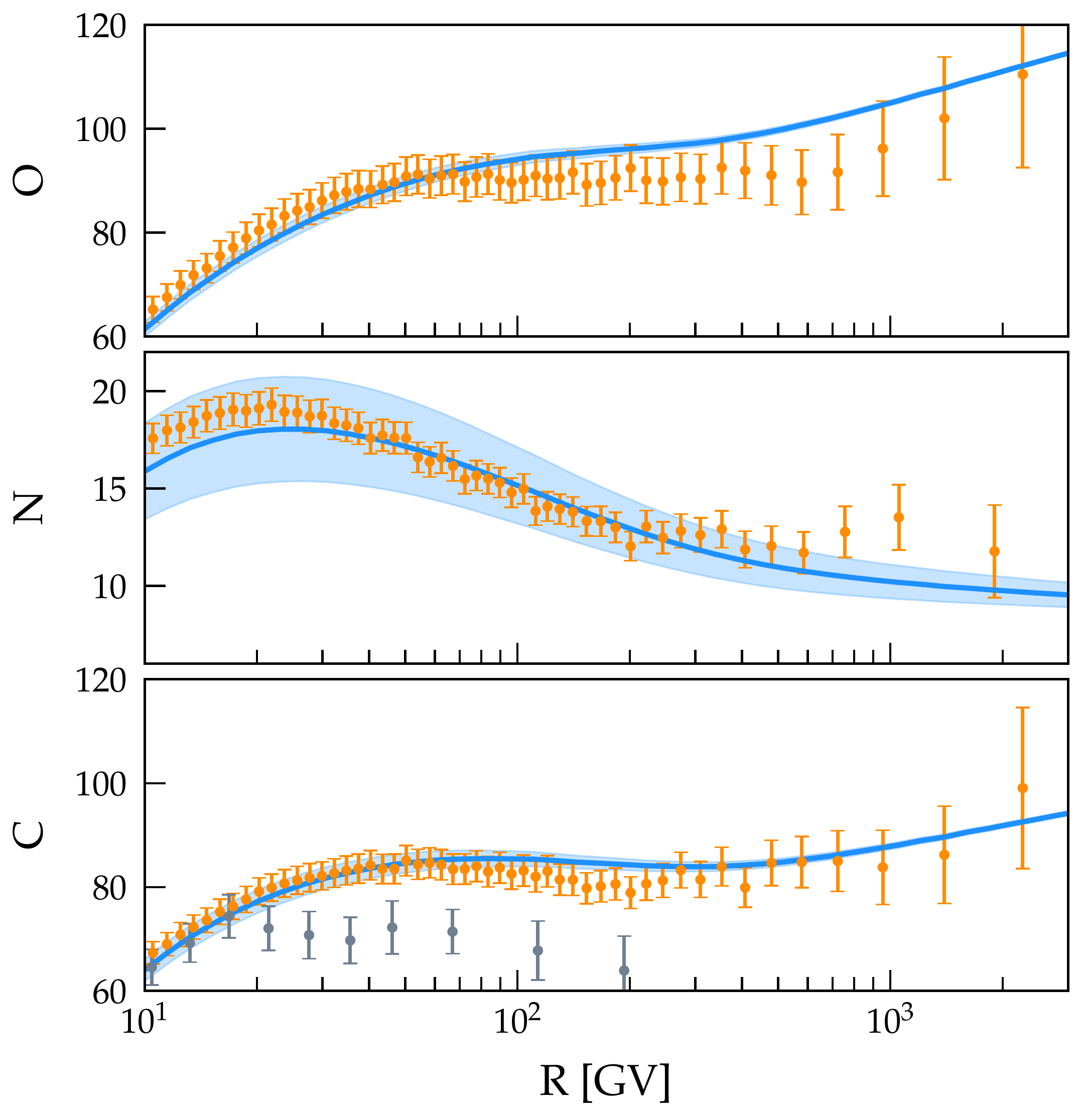}
\hspace{\stretch{1}}
\includegraphics[width=0.98\columnwidth]{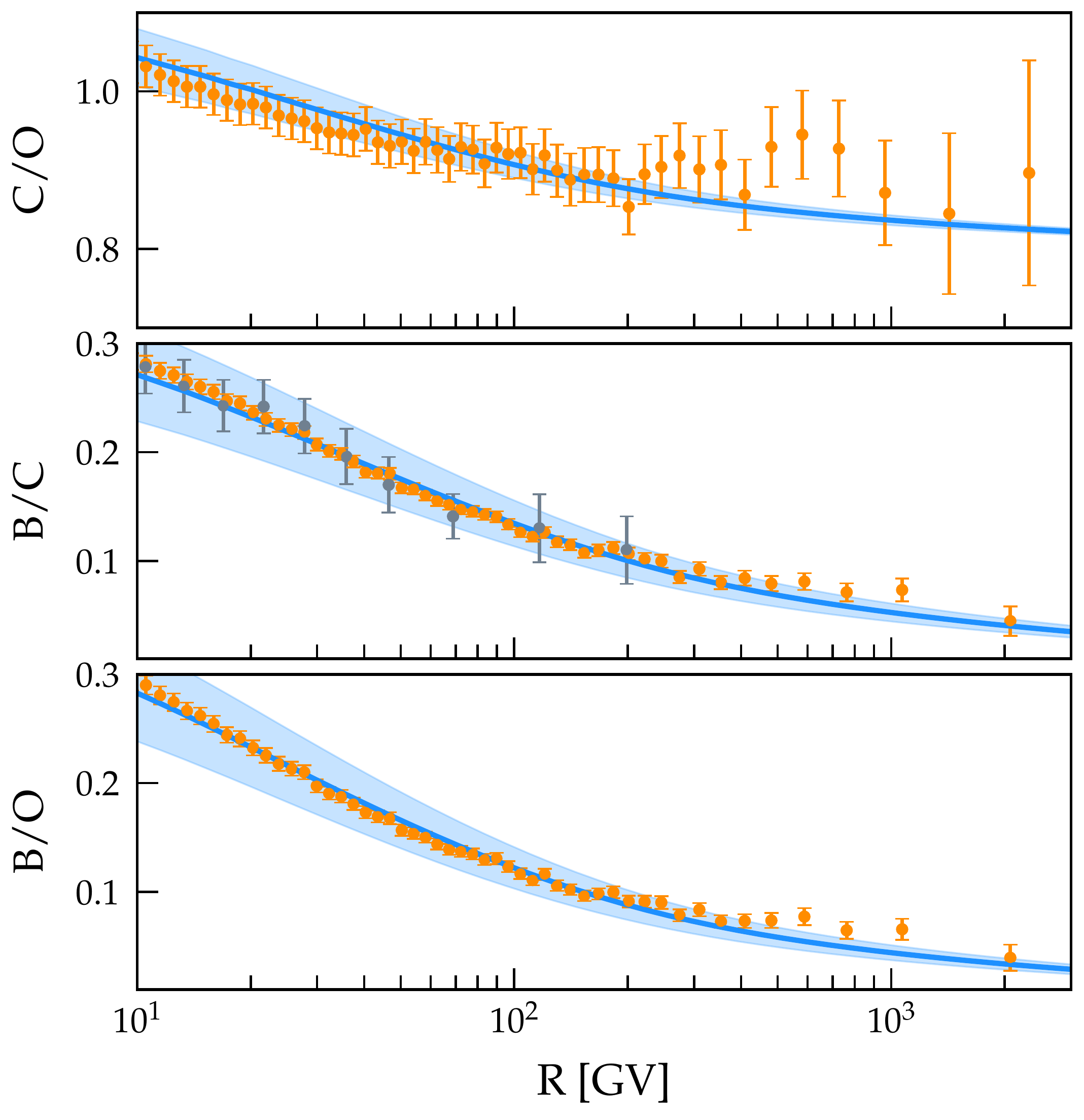}
\end{center}
\caption{The same as in Fig.~\ref{fig:bestfit} for the case in which the fit includes He nuclei as discussed in~\S~\ref{sec:B}.}
\label{fig:bestfit_modelb}
\end{figure*}

The measured He spectrum cannot be described by the results of our calculations, unless He is assumed to be injected with a spectrum that is harder than that of heavier nuclei by $\sim 0.05$. This conclusion appears to be rather robust: in the bottom panel of Fig.~\ref{fig:bestfit_HHe} one can see that the uncertainty in the He flux deriving {taking into account the} poorly known partial cross sections (shaded area) remains far from the data. Interestingly the data would be well described if the contribution of $^{3}$He were neglected (dashed line), which would clearly be at odds with what we know about the cross section of $^{3}$He production and $^{4}$He spallation~\cite{Coste2012}.

The disappointing conclusion of this first part of  our work is that a self-consistent description of the nuclei heavier than He and the C/O, B/C and B/O ratios can be obtained only to the extent that we accept a different injection spectrum for protons, He and heavier nuclei, a picture that would not be easy to justify on the basis of known models of acceleration and transport. 

Interestingly, the diffusion coefficient behaviour at energies below the break is very close to the one expected in the scenario in which diffusion is dominated by the self-generation of magnetic turbulence due to streaming instability~\cite{Aloisio2013}.

\subsection{Fit to nuclei including He}
\label{sec:B}

Inspired by the theoretical expectation to have the same injection spectrum for all nuclei, here we include He in the best fit to the data, with the same injection spectrum as heavier nuclei. The minimization procedure returns the following values of the parameters of our problem: $u = 6.6$~km/s, $D_0 = 1.1 \times 10^{28}$~cm$^2$/s, $\delta = 0.63$ and $\gamma = 4.23$, very similar to the previous case, with the exception of the harder injection spectrum. As one could expect, the measured He spectrum is very well described but some tension appears for oxygen. 

The left panel of Fig.~\ref{fig:bestfit_modelb} shows our results for the spectra of C, N and O, while the right panel shows the C/O, B/O and B/O ratios. The spectra of C and N are well described in this second case, but the oxygen spectrum appears to have a trend at odds with data, especially at $R\gtrsim 100$ GV, although a quantitative test of the goodness of the fit only shows a mildly higher $\chi^{2}$. The C/O, B/C and B/O ratios also appear to be well described.

\begin{figure}
\begin{center}
\includegraphics[width=0.98\columnwidth]{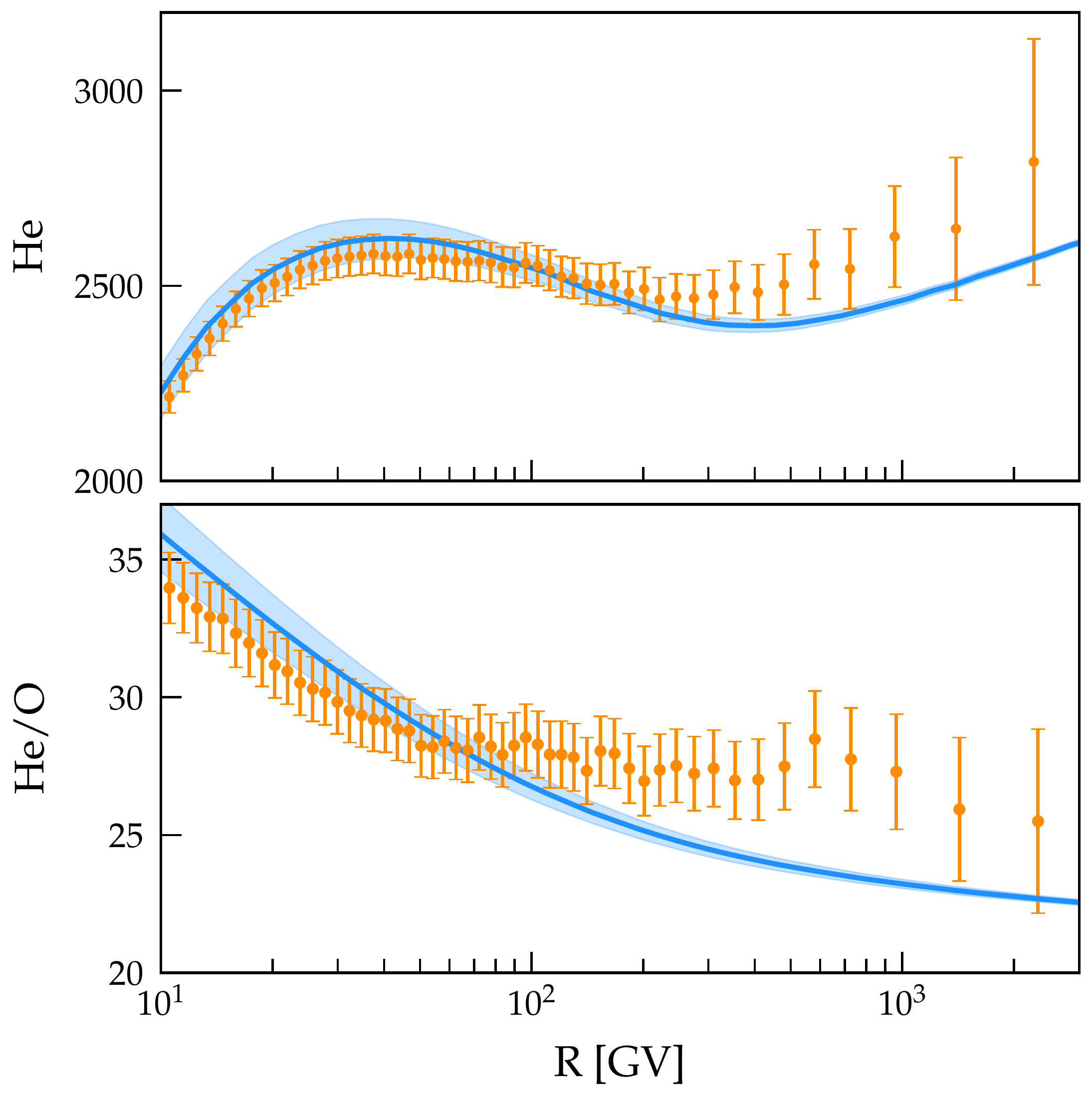}
\end{center}
\caption{The helium spectrum and the He/O ratio when all fluxes from He to O are injected with the same slope as in \S~\ref{sec:B}. Flux units as in Fig.~\ref{fig:bestfit}.}
\label{fig:He_modelb}
\end{figure}

The oddness of the O spectrum is made more clear in Fig.~\ref{fig:He_modelb} which shows the He spectrum (top panel) and the He/O ratio (lower panel). The predicted excess in the He/O ratio at low energies remains compatible with the data once the uncertainties in the cross section (shaded area) is included. On the other hand, the energy dependence of the ratio at high energies is quite different from that shown by the data, which appear to require a smaller grammage (less effective O spallation). The latter would however not be compatible with the grammage inferred from the B/C ratio. 

\begin{figure}
\begin{center}
\includegraphics[width=0.98\columnwidth]{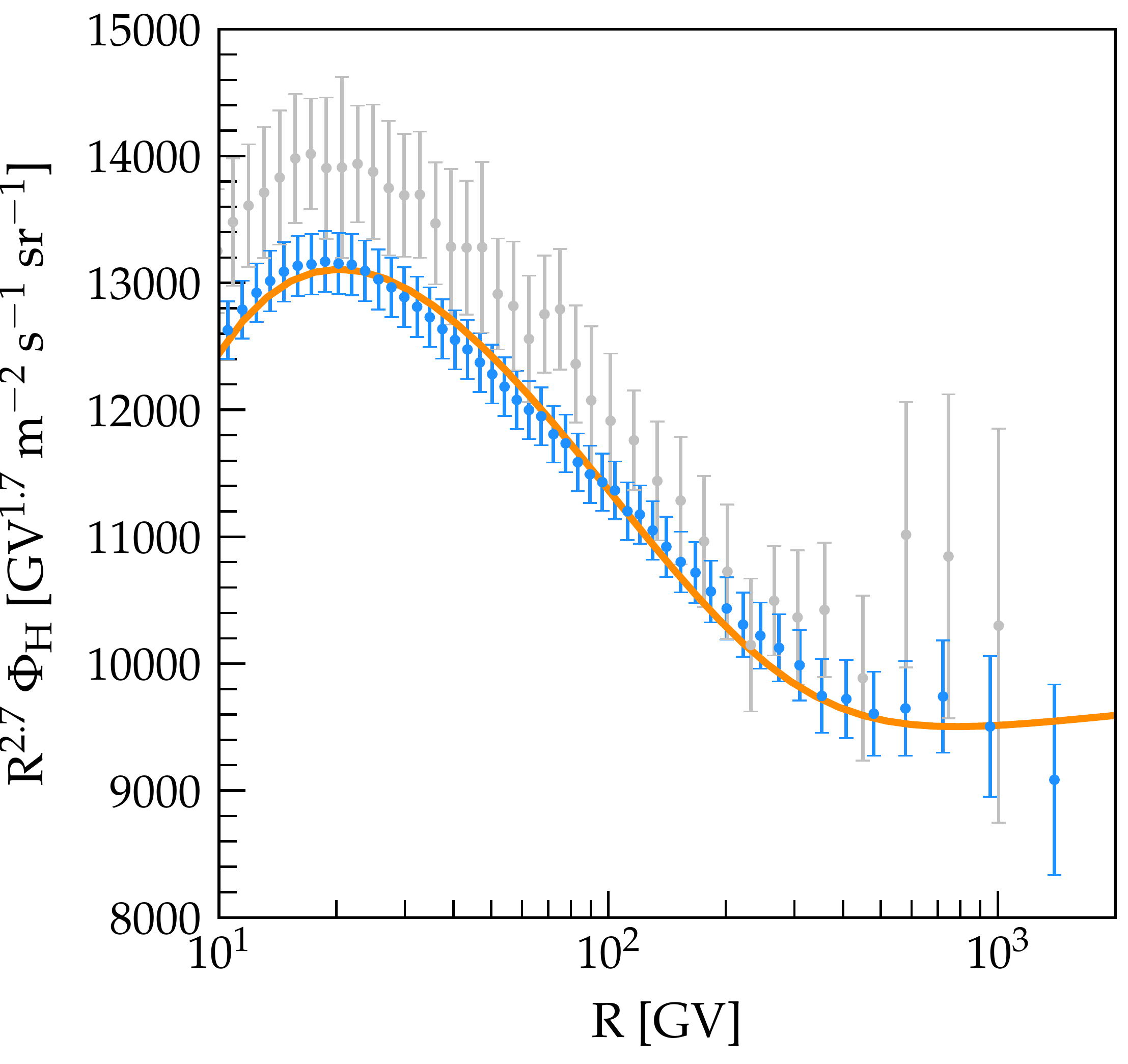}
\end{center}
\caption{The proton spectrum with an injection slope $\gamma = 4.31$ and efficiency $\alpha_{\rm H} = 4$\% is compared with AMS-02 (blue) and PAMELA (gray) data.}
\label{fig:H_modelb}
\end{figure}

A good fit to the proton spectrum requires an injection spectrum that is harder than that of nuclei by about $0.08$, namely $\gamma_{H}=4.31$. This is shown in Fig.~\ref{fig:H_modelb} together with the AMS-02 and PAMELA data. 

\subsection{Source grammage}
\label{sec:C}

CR interactions suffered during the acceleration process lead to the production of secondary nuclei that may appear in the observed secondary-to-primary ratios. The grammage traversed by CRs inside the sources is referred to as {\it source grammage} and, for typical values of the parameters of a SNR, assuming that they are the sources of CRs, the source grammage can be estimated to be $X_s \sim 0.2$~gr~cm$^{-2}$~\cite{Aloisio2015}. Being related to the residence time of CRs inside accelerators, $X_{s}$ is expected to be roughly energy independent, although some exceptions can be easily envisioned.

Since this additional contribution to the grammage is accumulated on a much shorter timescale than galactic escape, it is expected to manifest itself as a source term to  secondary species with a primarylike injection slope and normalization proportional to $X_s$:
\begin{equation}
Q_X^\alpha = 2 h_d \delta(z) X_s \sum_{\alpha' > \alpha} \frac{\sigma_{\alpha' \rightarrow \alpha}}{m} q_{0,\alpha'}(p) ,
\end{equation}
where the sum is made on species $\alpha'$ heavier than $\alpha$.

\begin{figure}
\begin{center}
\includegraphics[width=0.98\columnwidth]{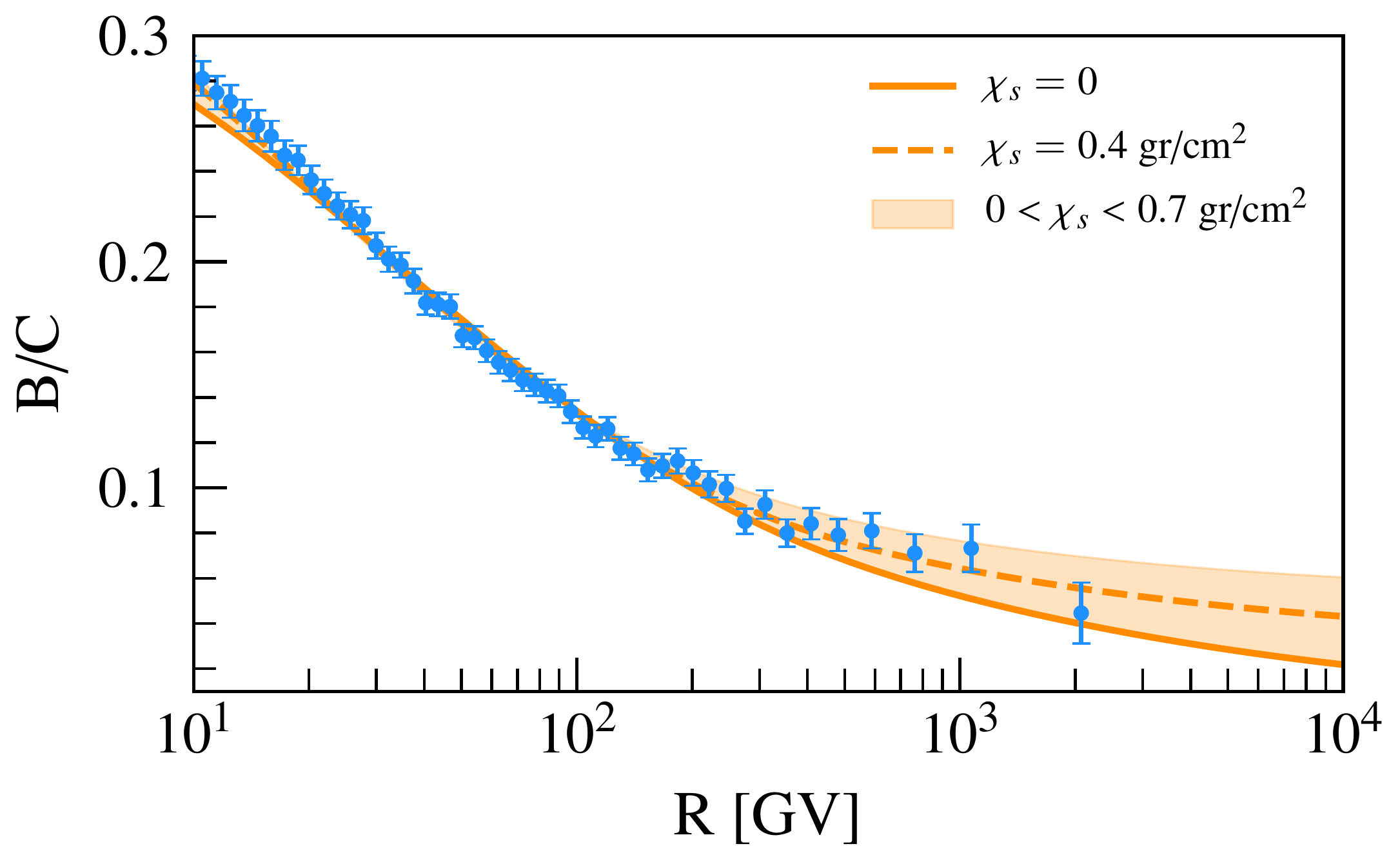}
\includegraphics[width=0.94\columnwidth]{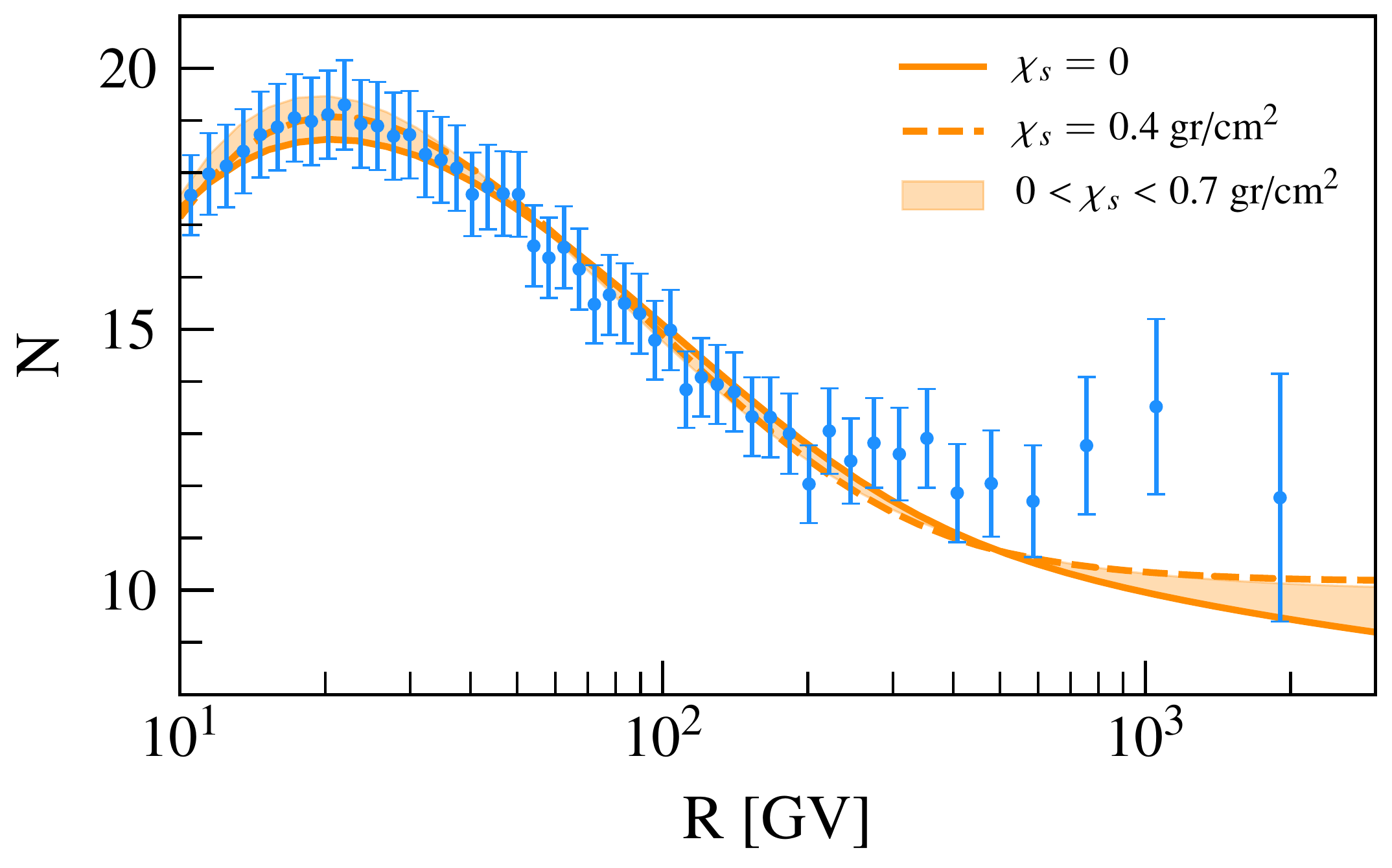}
\end{center}
\caption{The best-fit obtained with/without grammage at the source (\S~\ref{sec:C}) is shown for B/C (top panel) and nitrogen flux (bottomo panel) with a dashed/solid line. The shaded area represents the B/C ratio and the N flux for the range $0 < X_s < 0.7$~gr/cm$^2$.}
\label{fig:BC_Xs}
\end{figure}

The dashed area in Fig.~\ref{fig:BC_Xs} shows the result of adding the grammage accumulated by CRs inside the source to the one due to propagation in the Galaxy. For each value of the source grammage $X_s$ we fit the data to obtain a combination of parameter compatible with B/C and C and O spectra. We find that the range for $X_s$ that allows us to reproduce the B/C is $0 < X_s < 0.7$~gr/cm$^2$, while the best-fit is obtained for $X_s \sim 0.4$~gr/cm$^2$ (see Fig.~\ref{fig:BC_Xs}). The transport parameters that are more susceptible to the effects of the source grammage are $D_0$ and $\delta$. We found that, when $X_s = 0.4$~gr/cm$^2$, their best-fit values are $D_0 = 1.2 \times 10^{28}$~cm$^2$/s and $\delta = 0.68$. Correspondingly the injection slope of nuclei takes now the value $\gamma = 4.22$.

Being almost secondary at low energy, the nitrogen flux is also affected by the presence of source grammage. In the same figure, the nitrogen flux is shown for the case with and without the source grammage. In particular we consistently verified that nitrogen is well reproduced if its injection efficiency is reduced by $\sim 10$\%.

\begin{figure}
\centering
\includegraphics[width=0.94\columnwidth]{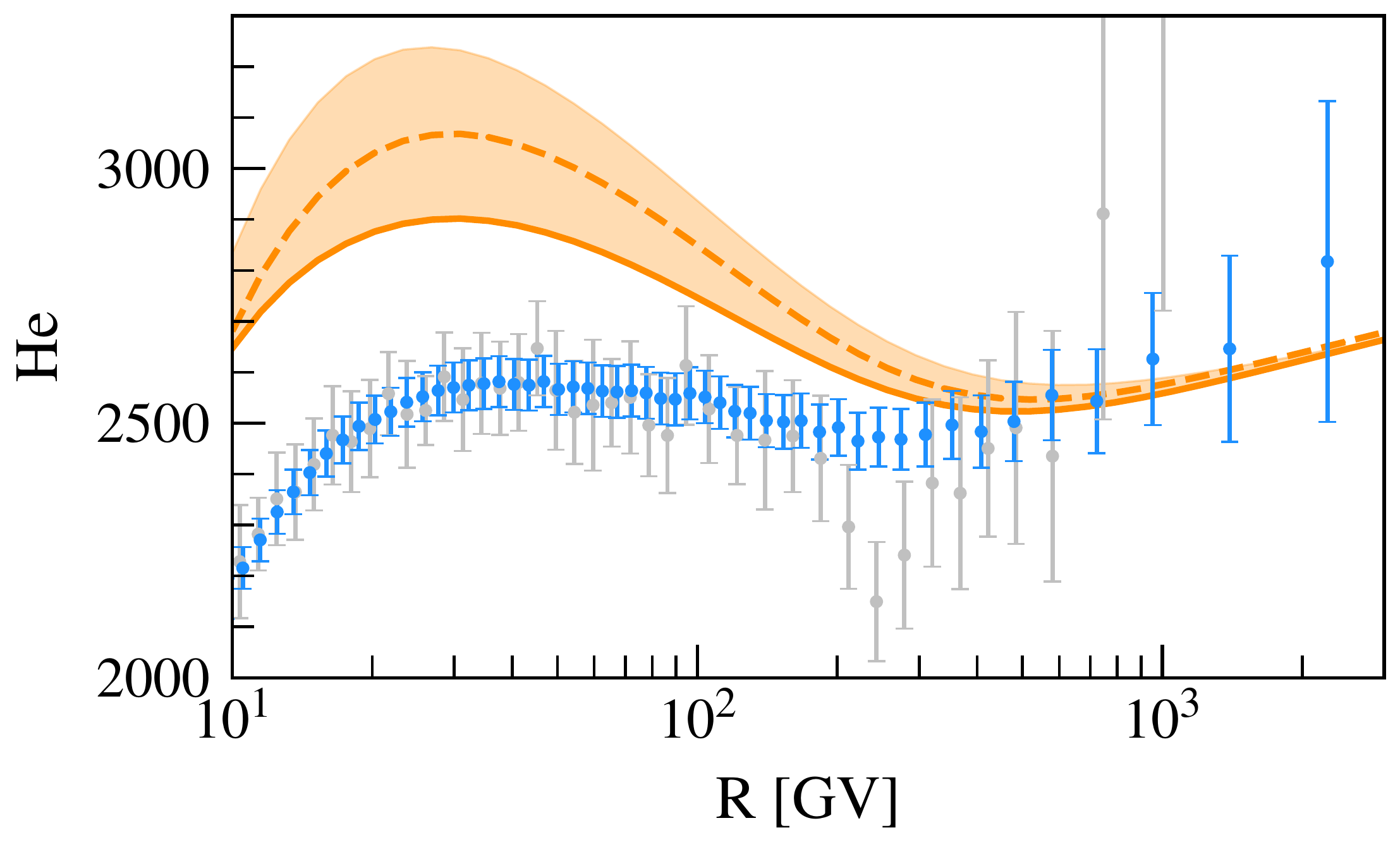}
\caption{Spectrum of helium nuclei. Lines are labeled as in Fig.~\ref{fig:BC_Xs}.}
\label{fig:He_Xs}
\end{figure}

On the other hand, we found no way to improve the agreement with the He flux: the source grammage leads to an enhanced production of secondary $^{3}$He, and in order to fit the AMS-02 data we are forced to require an even harder injection spectrum, $\gamma_{\rm He} = 4.15$, for $^{4}$He (see Fig.~\ref{fig:He_Xs}).

\section{Conclusions}
\label{sec:conclude}

The high precision of the AMS-02 data have changed the field of CR physics: on one hand they allowed us to put on solid grounds and to study in detail some well-known anomalies, such as the rise with energy of the positron fraction. On the other, this high precision opened the way to testing the solidity of the pillars of our standard model of the origin of CRs in an unprecedented way, just using the spectra of primary and secondary nuclei as measured by AMS-02, also including some important but often forgotten effects such as diffusive shock reacceleration~\cite{Blasi2017,Bresci2019}. 

In this article we focused our attention to the latter line of thought. We first proceeded to fit the AMS-02 data on the spectra of primary nuclei heavier than helium and the ratios of boron to carbon and boron to oxygen fluxes, as well as the C/O. All of these ratios provide information about the grammage traversed by CRs during their journey through the Galaxy and demanding self-consistency of these pieces of information is a basic request. The only physical assumption that we adopted here is that the change of slope in the spectrum of protons that has been observed by PAMELA and by AMS-02 independently originates from a change of regime in the diffusion of CRs~\cite{Genolini2017}, rather than being due to some subtle effects of the acceleration processes or some proximity effects of the sources. In this context, the proton spectrum is only used to characterize the change of slope in the energy dependence of the diffusion coefficient, while the absolute slope of the latter is directly obtained from a combined fit to the spectra of primary and secondary nuclei.

The fit to the spectra of nuclei heavier than He led to a very good description of the data, corresponding to the following values of the free parameters of the problem: $u = 7$~km/s, $D_0 = 1.1 \times 10^{28}$~cm$^2$/s, $\delta = 0.63$ and $\gamma = 4.26$, where the slope of the injection spectrum $\gamma$ is assumed to be the same for all such nuclei.  
However, this conclusion leads to a bad description of the spectra of protons and He nuclei, which instead require respectively a softer and harder injection by about $\sim 0.05$ with respect to nuclei. In other words, not only is $^{4}$He expected to have an injection spectrum harder than that of hydrogen, but it is also required that nuclei are injected with a softer spectrum than $^{4}$He. Interestingly this finding is at odds even with models of the difference between protons and helium injection based on the different $A/Z$ ratio at shocks~\cite{Malkov2012,Hanusch2019}.

We carried out the calculation described above keeping track of the experimental uncertainty in the spallation cross sections: we clearly showed that the error bars in the measured fluxes of primary and secondary nuclei are already smaller than the uncertainties induced by the cross sections. Hence, improving further the systematics of the measurements or accumulating more statistics would not lead to a dramatic improvement in our understanding of the origin of CRs. A possible exception to this conclusion might apply to the highest energy bins ($\gtrsim 100$ GV), where the statistics of events remains the limiting factor. 

The conclusion that data seem to be best described by adopting three different injection spectra for protons, helium and heavier nuclei is clearly unsatisfactory and at odds with the rigidity dependence expected for CR transport inside and outside accelerators. Hence we repeated the calculations illustrated above by imposing a fit to the helium spectrum, in addition to the fluxes of C, N, O and the secondary nuclei and requiring helium to be injected with the same spectrum as the heavier elements. The fit that we obtained is clearly worse than in the previous case, but still acceptable from the statistical point of view. Yet, the spectrum of oxygen is described rather badly with a clear excess at high energy, that also reflects in a bad description of the He/O ratio.  Interestingly, the oxygen flux seems to demand a smaller grammage that needed to fit the B/C ratio.

As a third and final part of this calculation, following some previous literature \cite{Aloisio2015}, we introduced an additional grammage that is expected to mimic the spallation reactions inside the acceleration region. Such source grammage $X_{s}$ is expected to be roughly energy independent, hence it leads to an additional production term of secondaries (such as boron) with the same spectrum as the injection, so that this contribution becomes important at high energies, although it does change the numerical fits to the data also at lower energies. The order of magnitude of this contribution is $X_{s}\sim 0.1-0.4~\rm g~cm^{-2}$. The idea is that the presence of the source grammage may require less boron production due to CR transport in the Galaxy, namely a larger value of $\delta$ and, as a consequence, a harder injection spectrum of nuclei. 
The formal fitting procedure does confirm this trend, but the additional production of $^{3}$He in the sources leads to requiring an even harder injection spectrum of $^{4}$He, thereby making the problem of the difference between He and other elements even more evident. 

As a conclusion, the high precision AMS-02 data, if taken at face value and within the realm of the standard picture of CR transport in the Galaxy, lead to the conclusion that the acceleration process and/or the process of escape of CRs from the sources are more complex than usually modeled: protons, He and heavier elements need to be injected into the ISM with different spectra, contrary to what expected in the common wisdom. This should be considered as a stimulus to investigate the physics of particle acceleration and escape from the sources more seriously than done so far. In alternative, the AMS-02 measurements might suggest some major modification of the paradigm of CR transport, as recently discussed in Refs.~\cite{Cowsik2016,Lipari2017}. %
The requirements of these models have been recently reviewed in Ref.~\cite{Amato2018}.

\begin{acknowledgments}
We acknowledge the use of the CRDB~\cite{Maurin2014} and ASI~\cite{DiFelice2017} databases for providing CR measurements.
C.E.~acknowledges the European Commission for support under the H2020-MSCA-IF-2016 action, Grant No.~751311 GRAPES 8211 Galactic cosmic RAy Propagation: An Extensive Study.
This work was partially funded through Grants ASI/INAF No.~2017-14-H.O
\end{acknowledgments}

\appendix

\section{Cross sections measurements and energy dependent fitting functions}
\label{sec:appendix}

In this Appendix, we provide updated fits to the most relevant isotopic production cross sections for Li, Be, and B. In fact, the major production channels of these nuclei are due to spallation reactions of CNO primary nuclei, summing up to $\gtrsim 80$\% of the total LiBeB production at 10 GeV/n~\cite{Evoli2018}.
In this context it is also relevant to consider the fragmentation of CNO nuclei into $^{11}$C, since this is a radioactive isotope that decays in $^{11}$B with an half-life (at rest) of $\sim$20 minutes and the production cross sections from CNO is as large as the one in $^{11}$B.

Measurements of these cross sections (mainly from Webber and coworkers in the 1990s) have been collected by the GALPROP collaboration and distributed within their code\footnote{In the file \url{isotope_cs.dat}}. A list of references to these data sets can be found in~\cite{Genolini2018}.
Additional measurements of the relevant isotopic cross sections have been obtained by the authors of Ref.~\cite{Evoli2018} by querying the EXFOR (Experimental Nuclear Reaction Data) database\footnote{\url{https://www-nds.iaea.org/exfor/exfor.htm}} which is an extensive database of nuclear reactions containing experimental data, their experimental information and source of uncertainties.

In order to parametrize the fragmentation cross section $\sigma$ of the nucleus $j$ to a lighter species $i$ on a hydrogen target as a function of the kinetic energy per nucleon $T$ we follow~\cite{Reinert2018}:
\begin{multline}\label{eq:fragmentation}
\sigma_{j+{\rm H}\,\rightarrow\, i} 
= \sigma_{0} \, \frac{\Gamma^2 \, (T-E_{\text{th}})^2}{(T^2-M^2)^2 +\Gamma^2 M^2} \\
+ \sigma_{1} \left(1 - \frac{E_{\text{th}}}{T}\right)^\xi \left(1 + \frac{\Delta}{1+(T_\text{h}/T)^2} \right) \,.
\end{multline}

In doing so, we assume that the kinetic energy per nucleon is conserved in the reaction.
Above the energy threshold $E_{\text{th}}$, the cross sections in Eq.~\ref{eq:fragmentation} show a resonance peak whose normalization, position and width is set by the parameters $\sigma_{0}$, $M$ and $\Gamma$. 
On top of the peak, this expression allows for a steady rise which continues up to $T\sim$~GeV with its smoothness is controlled by $\xi \ge 0$. 
Given that the peak is not visible in all the channels, we do not account for the peak in the fit, namely we assume $\sigma_0 = 0$, if adding this does not improve significantly the $\chi$-squared computed against the data. 
The threshold energy $E_{\text{th}}$ can be measured for each reaction. We retrieved the energy thresholds from the online database of the NNDC (National Nuclear Data Centre)\footnote{http://www.nndc.bnl.gov/qcalc/index.jsp} and based on the experimental results reported in~\cite{Wang2017}.  

We notice that at energies where we are interested here, $T \gtrsim 10$ GeV/n, spallation cross sections in Eq.~\ref{eq:fragmentation} approach the asymptotic value:
\begin{equation}
\lim_{T \gg {\rm GeV/n}} \sigma_{j+{\rm H}\,\rightarrow\,\text{i}} \rightarrow \sigma_{1} \, (1 + \Delta) 
\end{equation}
the existence of a plateau above few GeV/n is commonly assumed in the literature (see however~\cite{Genolini2017} for an attempt to consider a mild energy dependence on the partial cross sections) but still not fully assessed on experimental basis.
As in~\cite{Webber2003}, we allow for a slow change of the cross section around $T_\text{h}=2$~GeV controlled by the free parameter $\Delta$. This behavior has been introduced by Webber and collaborators to better reproduce the measurements.

To minimize the $\chi^2$ we use the MINUIT package\footnote{\url{https://github.com/jpivarski/pyminuit}} and we compute the confidence interval for the free parameters by means of the MIGRAD algorithm.
In Table~\ref{tab:fragcross} we report the best fit values and the 1-$\sigma$ uncertainty for the free parameters in our model given by Eq.~\ref{eq:fragmentation}.
In some cases, e.g., $\sigma_{{\rm N}^{14}+{\rm H}\,\rightarrow\, B}$, the measurements at our disposal are insufficient or at too low energy to perform a meaningful fit. 
For those channels we rely on the Webber parametrizations renormalized to the data  
whenever they are available (see discussion in \S~5 of~\cite{Evoli2018}).

The last column of Table~\ref{tab:fragcross} reports the reduced-$\chi^2$ computed with the best-fit parameter values.
For few channels, the $\chi^2$ is much larger than $\mathcal{O}(1)$. This is mainly due to the inconsistency between different data sets and/or to the under-estimation of the systematic errors associated with these measurements, in particular for experiments that took data before the 1990s.
In the same table we also report the relative uncertainty on the high-energy (plateau) value of the cross sections.
To compute this, we evaluate the minimum and maximum cross section value by combining the uncertainties of the parameters governing the high-energy behaviour of the model, namely $\sigma_1$, $\xi$, and $\Delta$.
We then estimate the relative uncertainty on the cross section normalization to be as large as 30\%.

\begin{turnpage}
\begin{figure}[htp]
\begin{center}
\includegraphics[width=0.31\textheight]{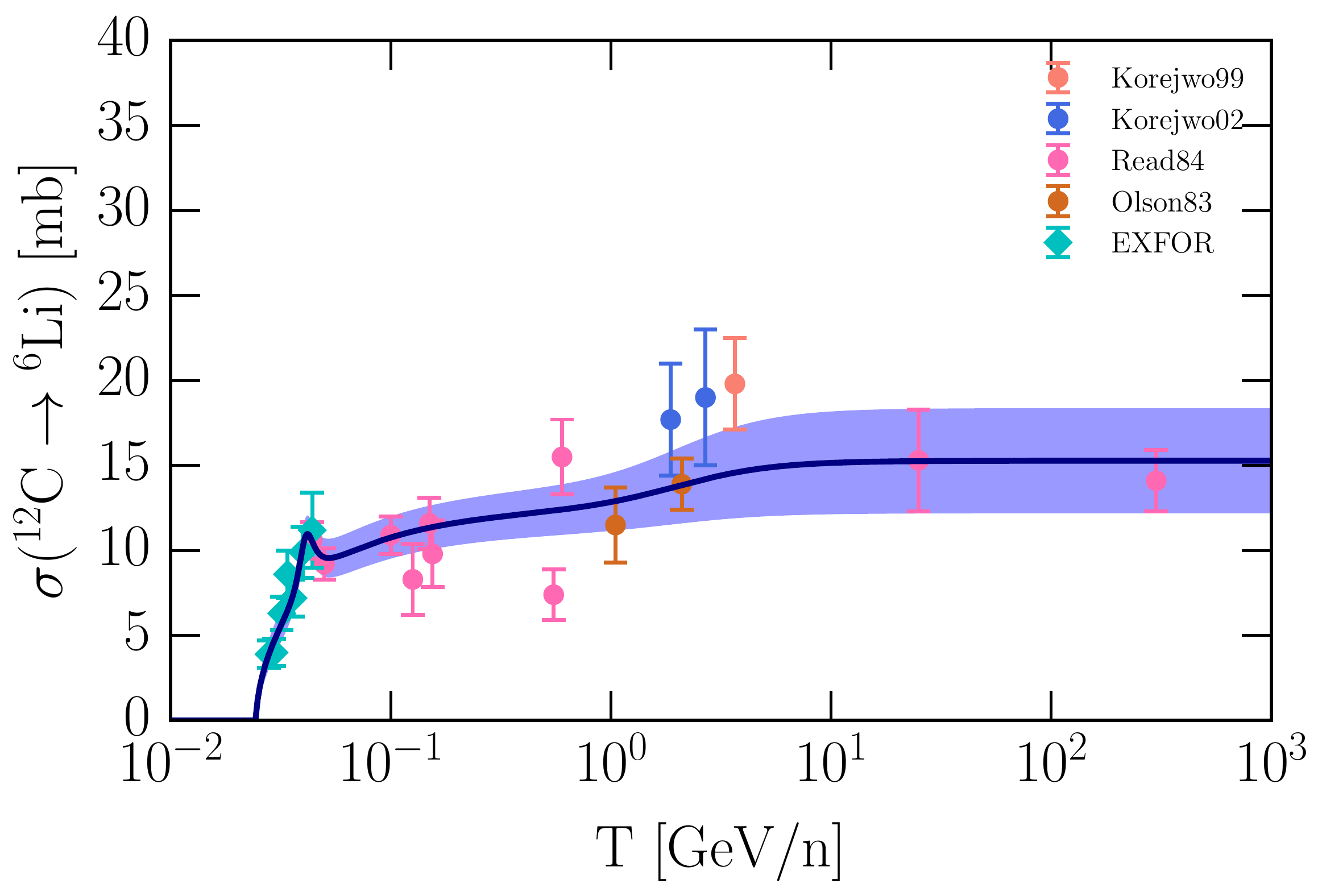}
\includegraphics[width=0.31\textheight]{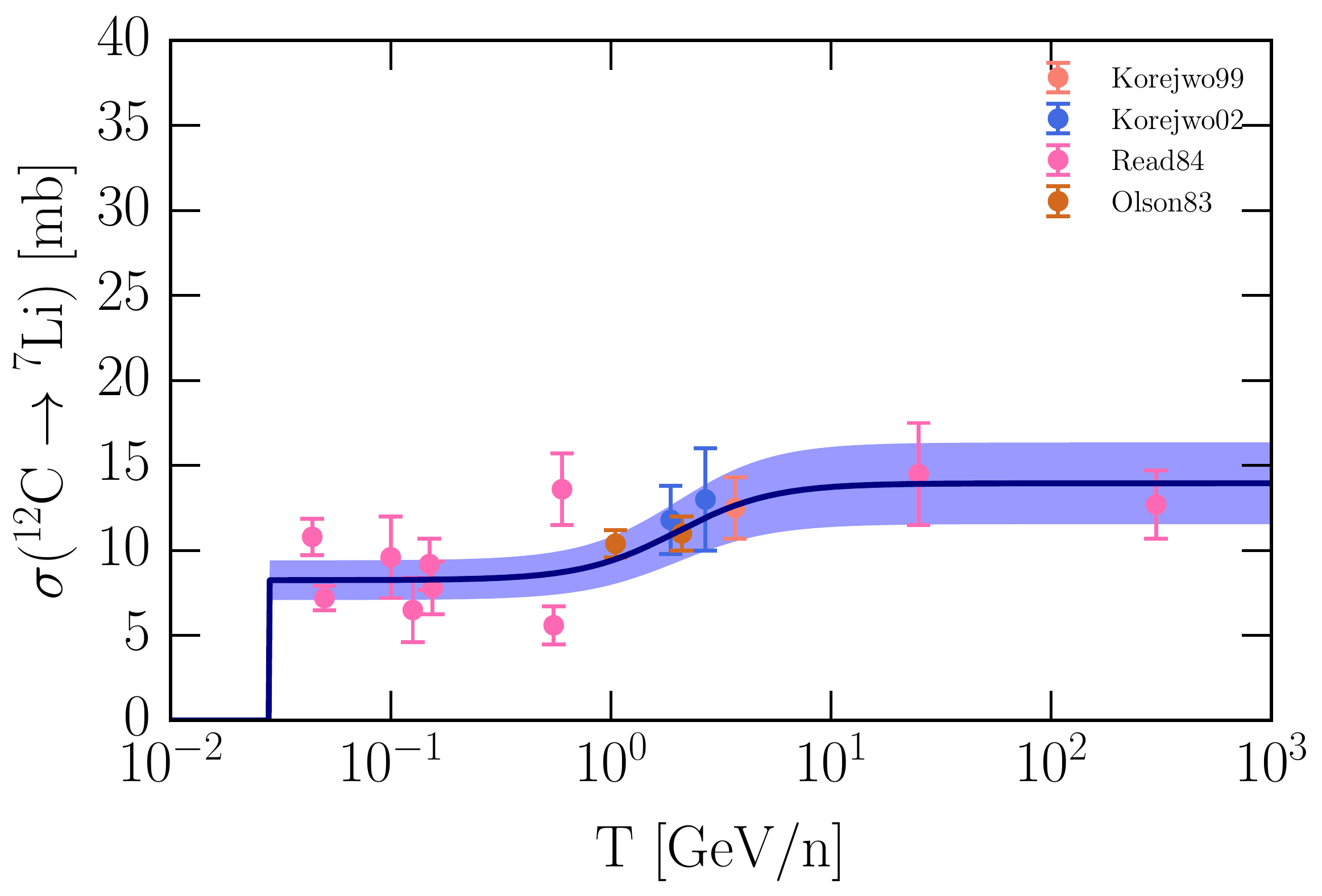} \\
\includegraphics[width=0.31\textheight]{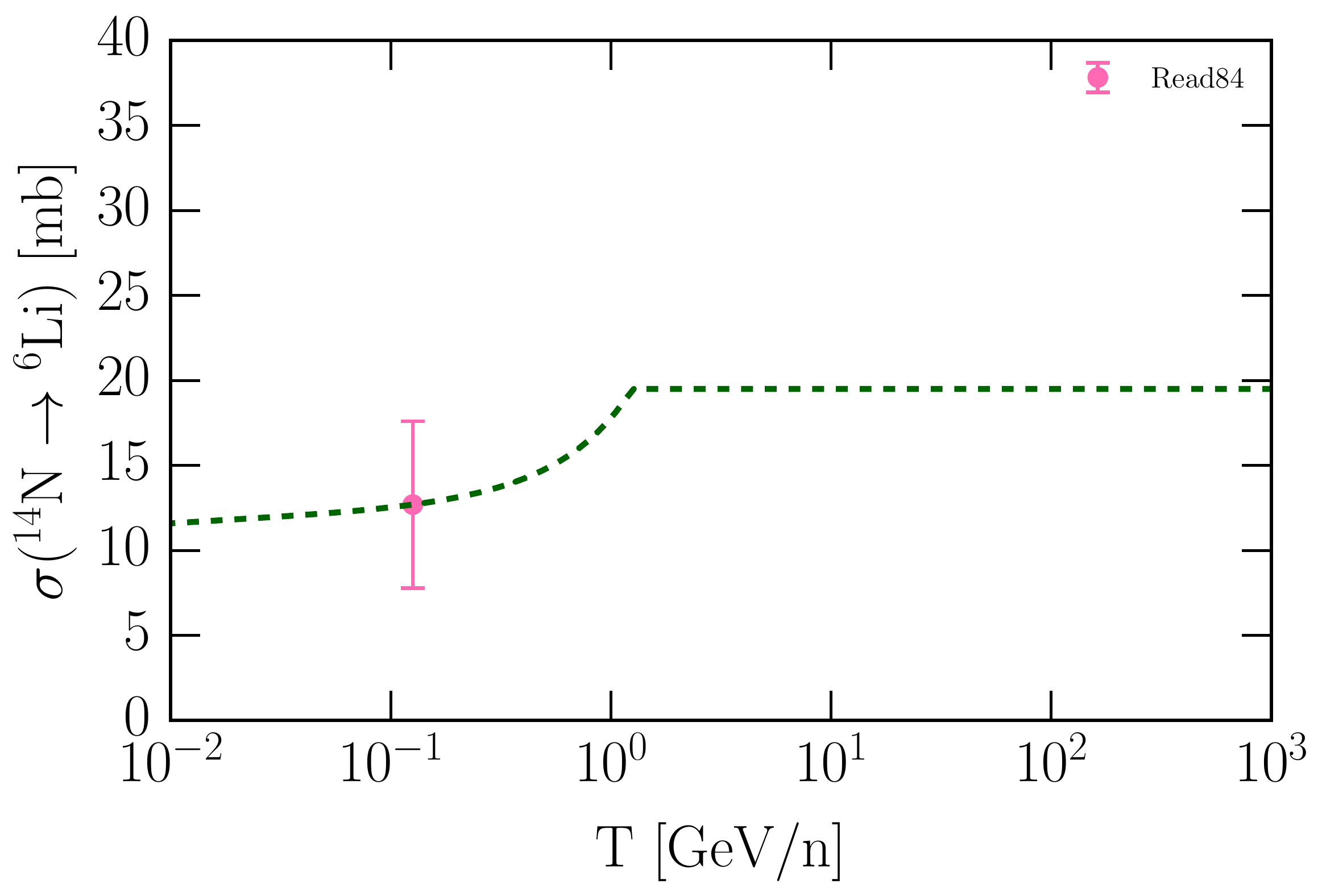}
\includegraphics[width=0.31\textheight]{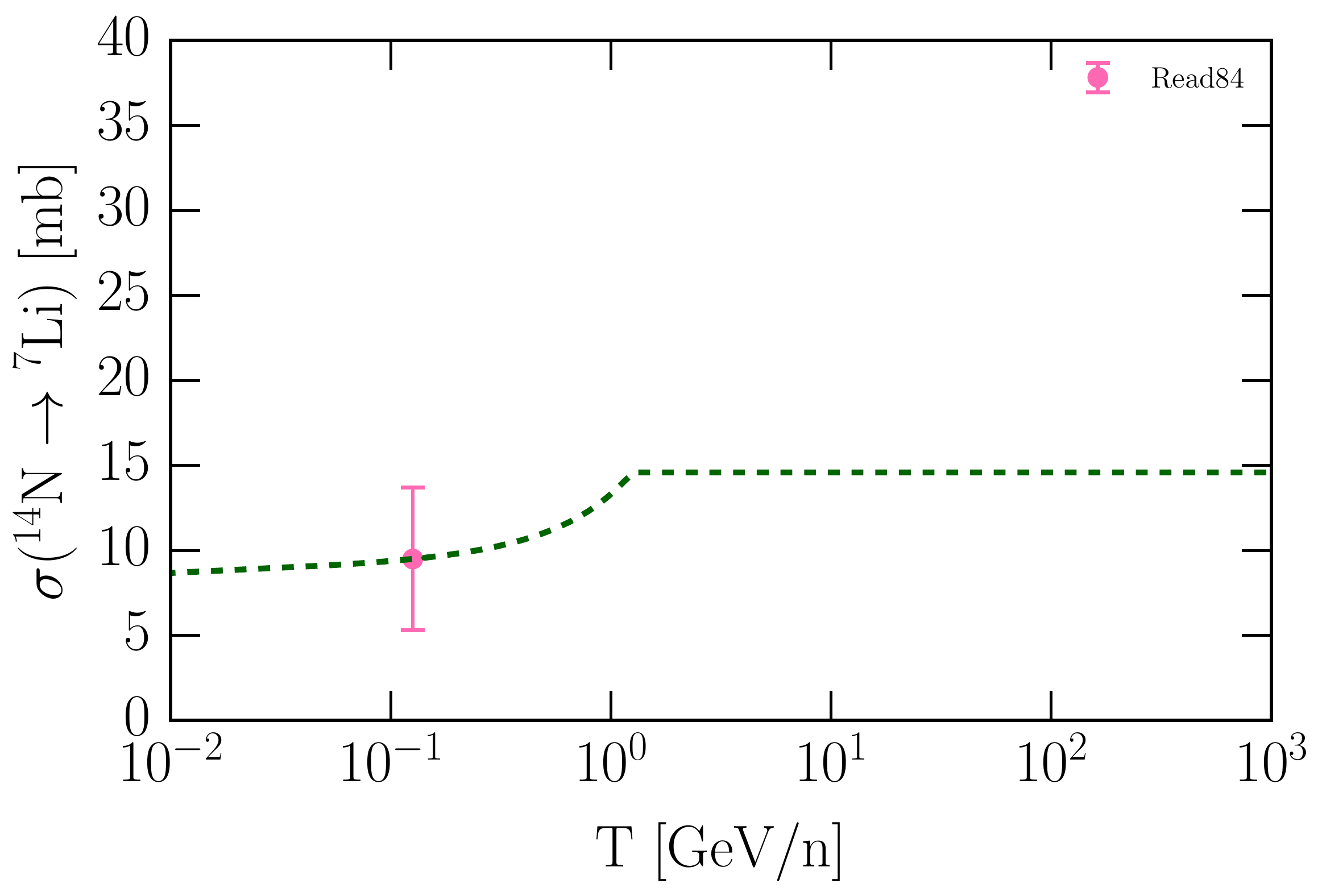} \\
\includegraphics[width=0.31\textheight]{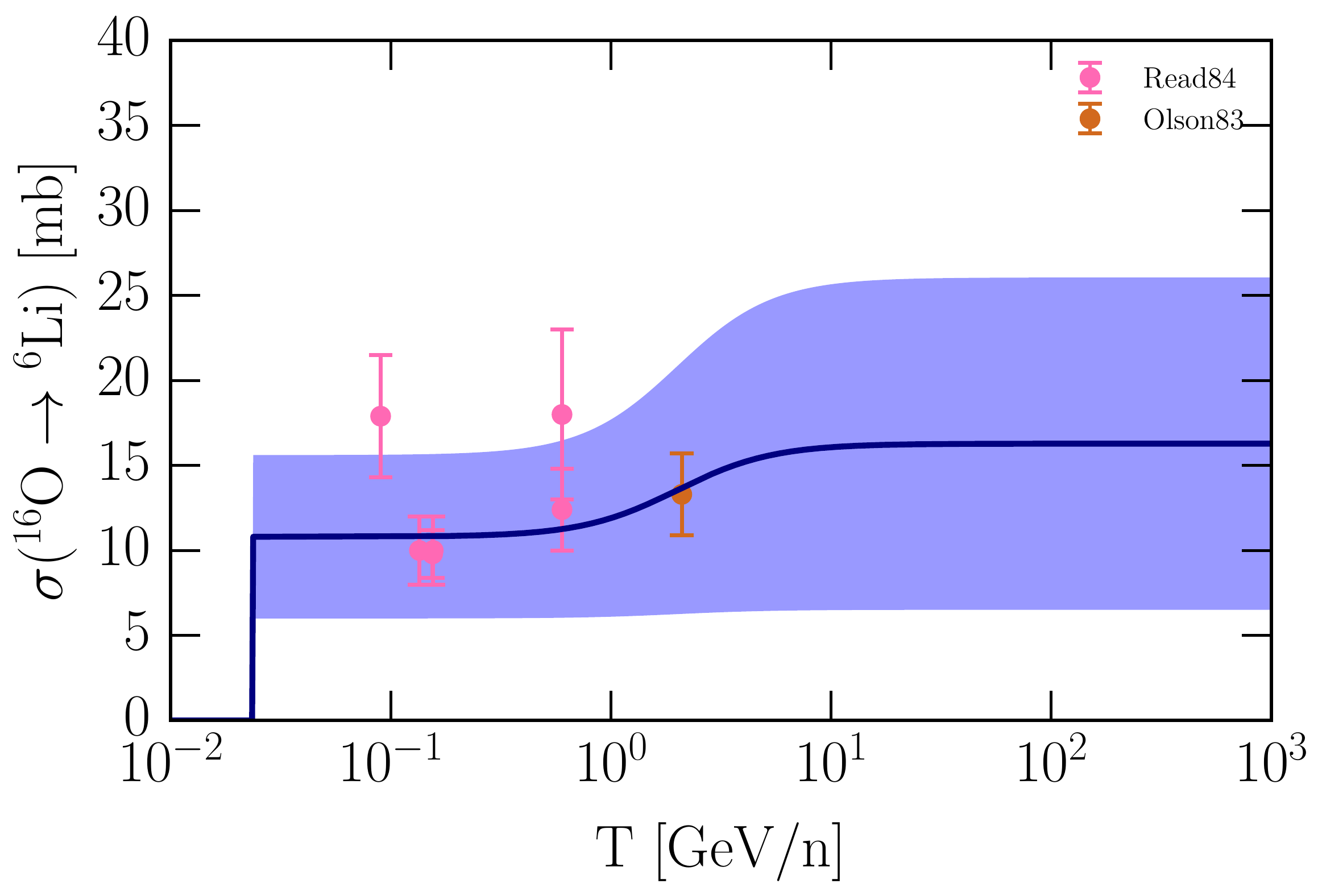}
\includegraphics[width=0.31\textheight]{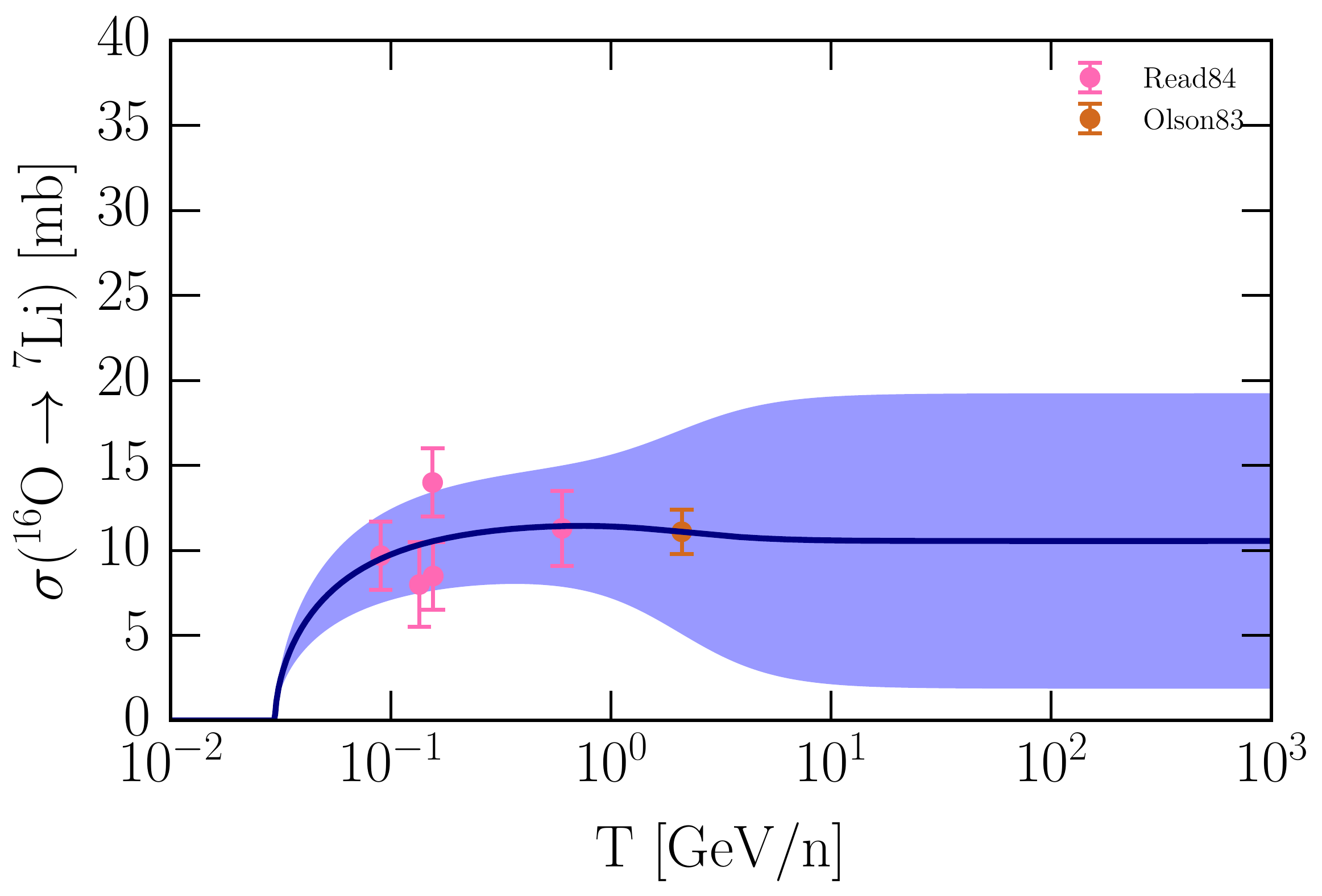} \\
\end{center}
\caption{Secondary production cross sections for Lithium isotopes. Solid line represents the fit based on Eq.~\ref{eq:fragmentation}. The shaded area extends between the minimum and maximum value allowed within 1-$\sigma$ of the parameters $\sigma_1$, $\xi$ and $\Delta$. Dashed green line shows the Webber parametrization normalized to the data.}
\label{fig:Lithium}
\end{figure}
\end{turnpage}

\begin{turnpage}
\begin{figure}[htp]
\begin{center}
\includegraphics[width=0.31\textheight]{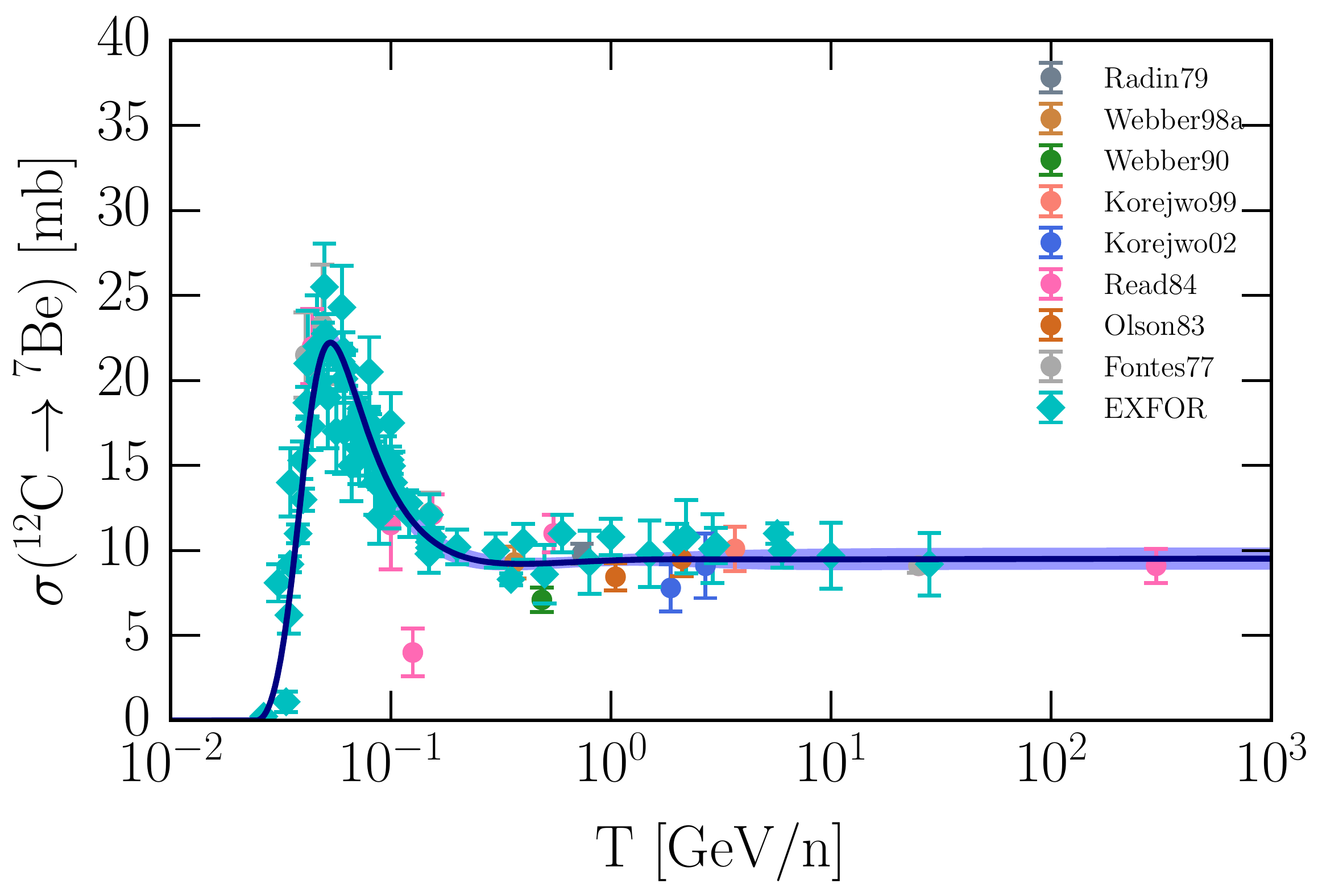}
\includegraphics[width=0.31\textheight]{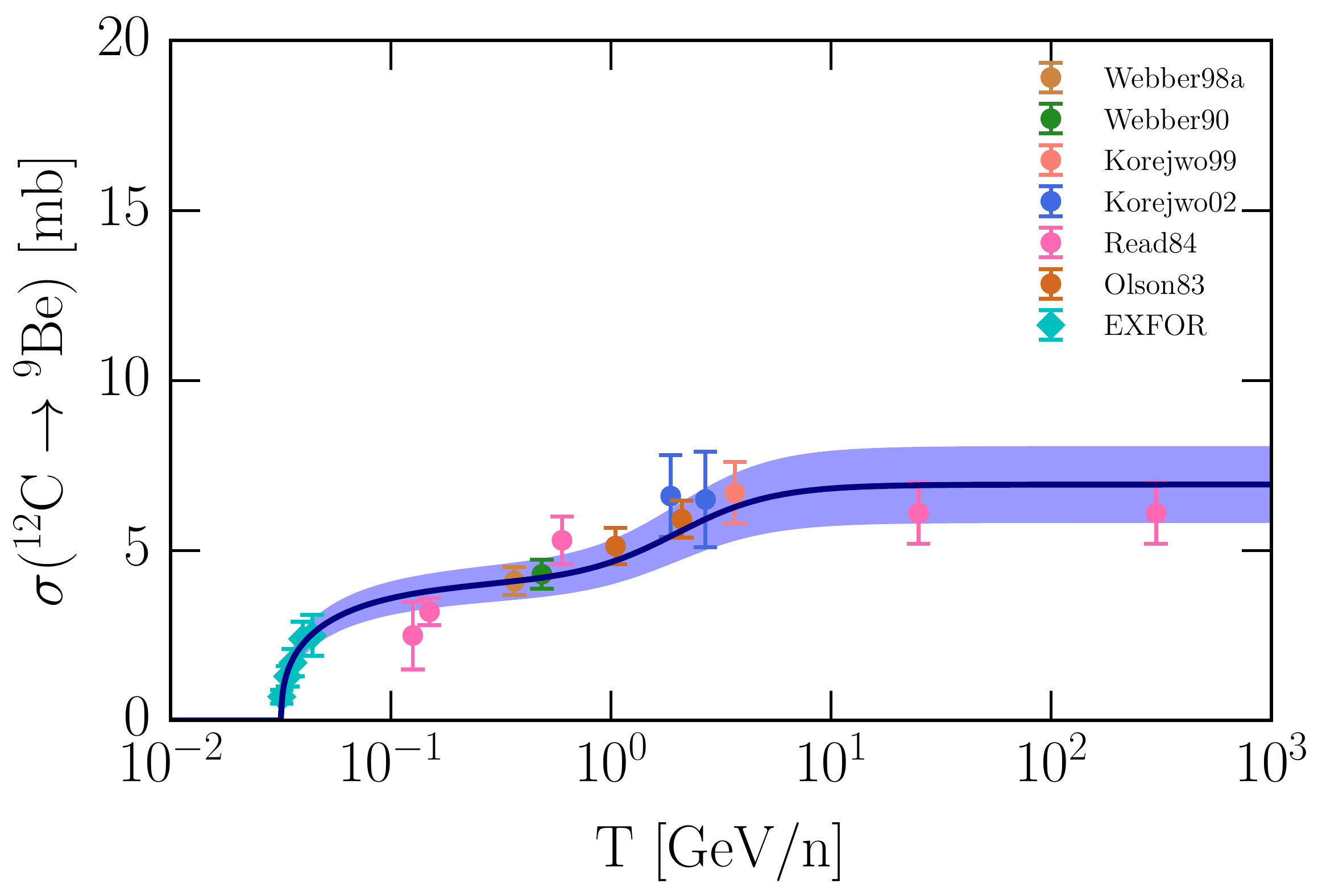} \includegraphics[width=0.31\textheight]{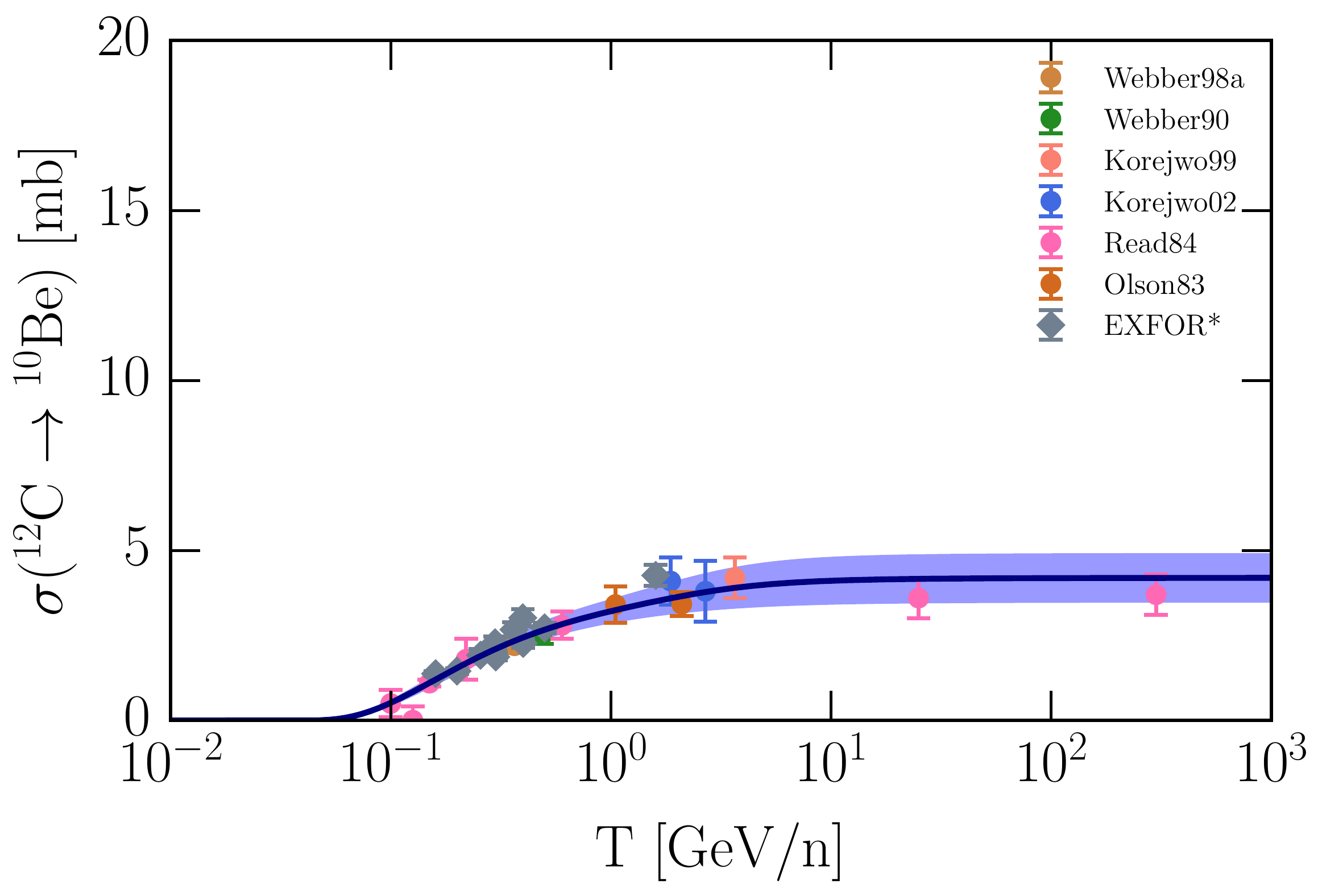} \\
\includegraphics[width=0.31\textheight]{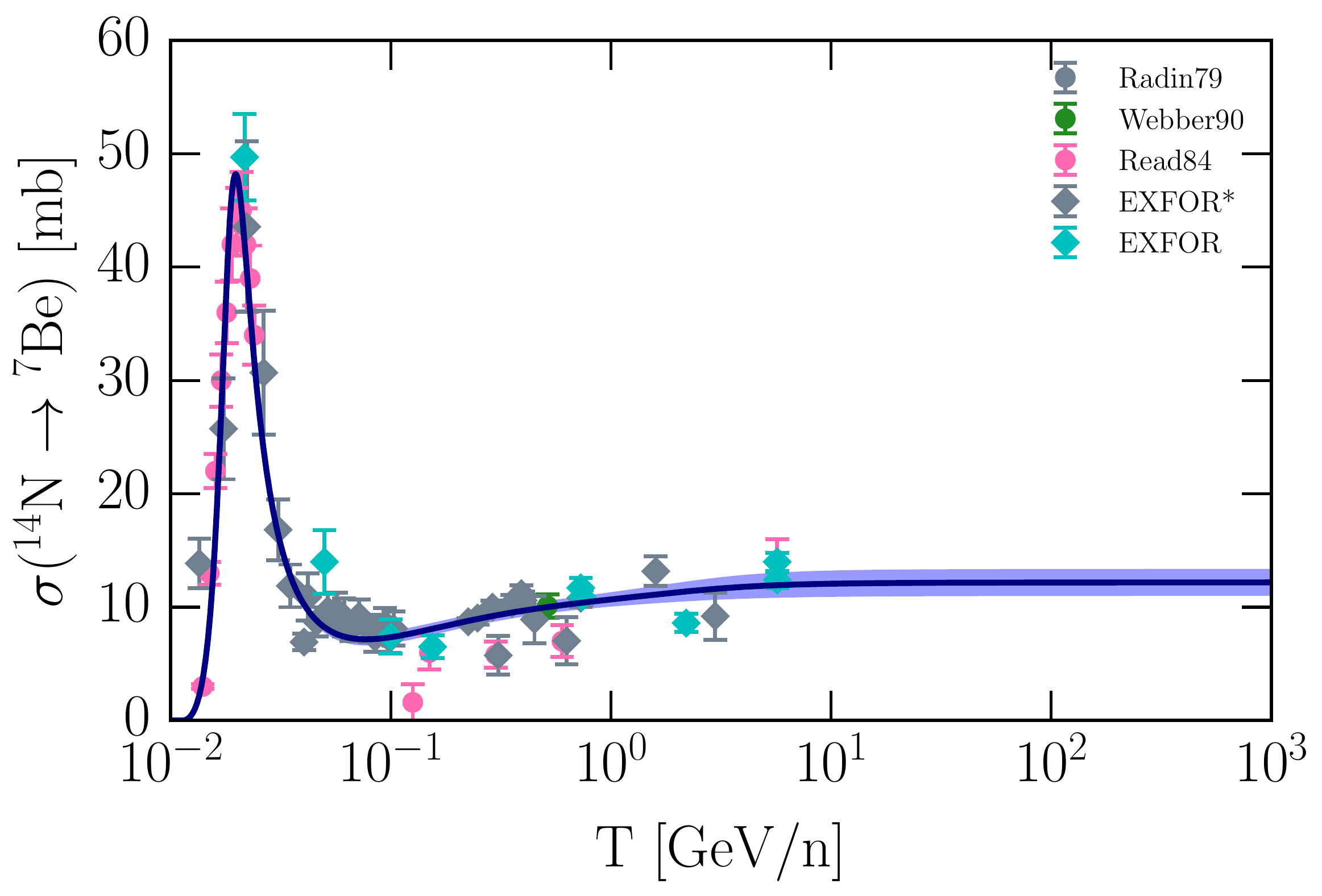}
\includegraphics[width=0.31\textheight]{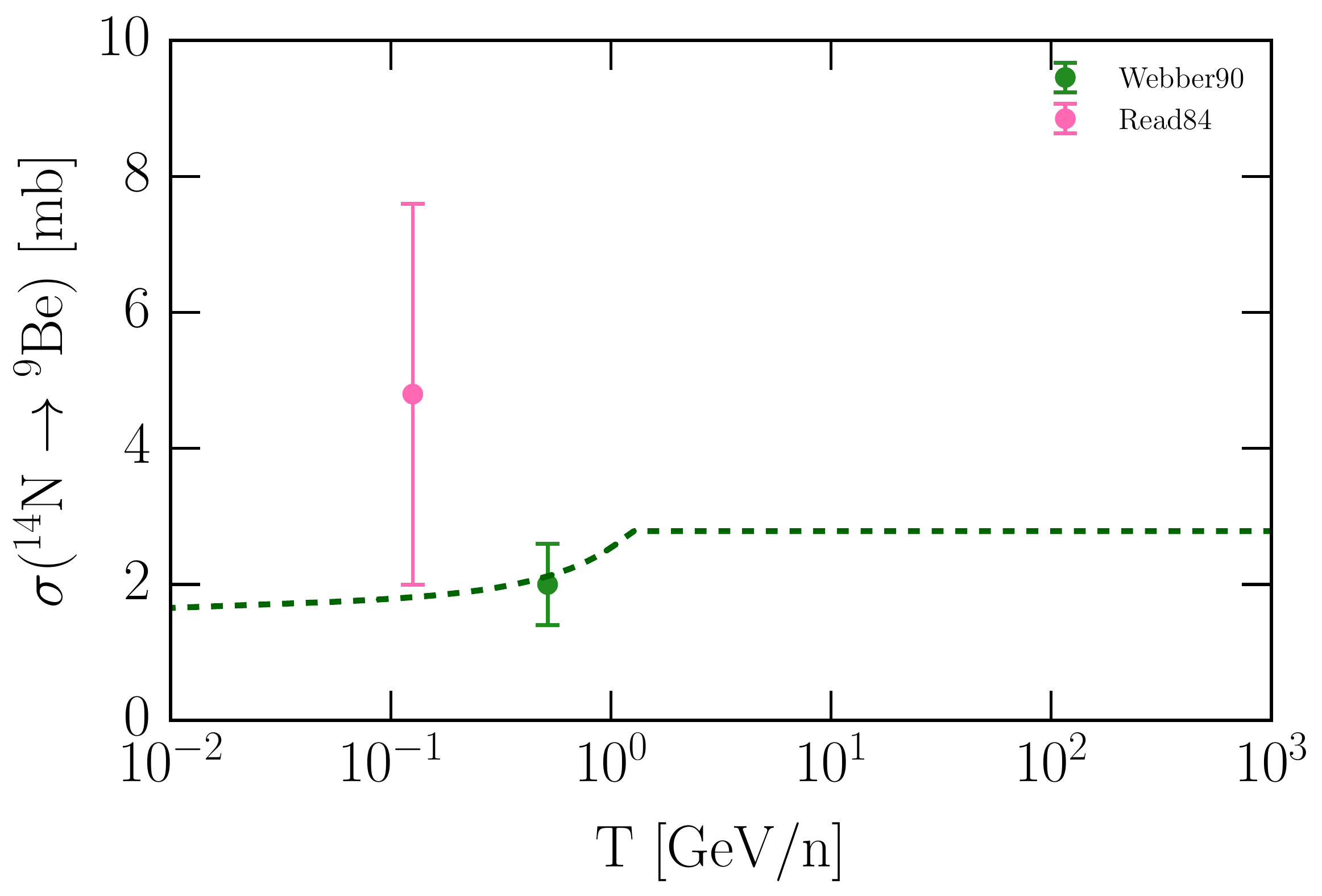} 
\includegraphics[width=0.31\textheight]{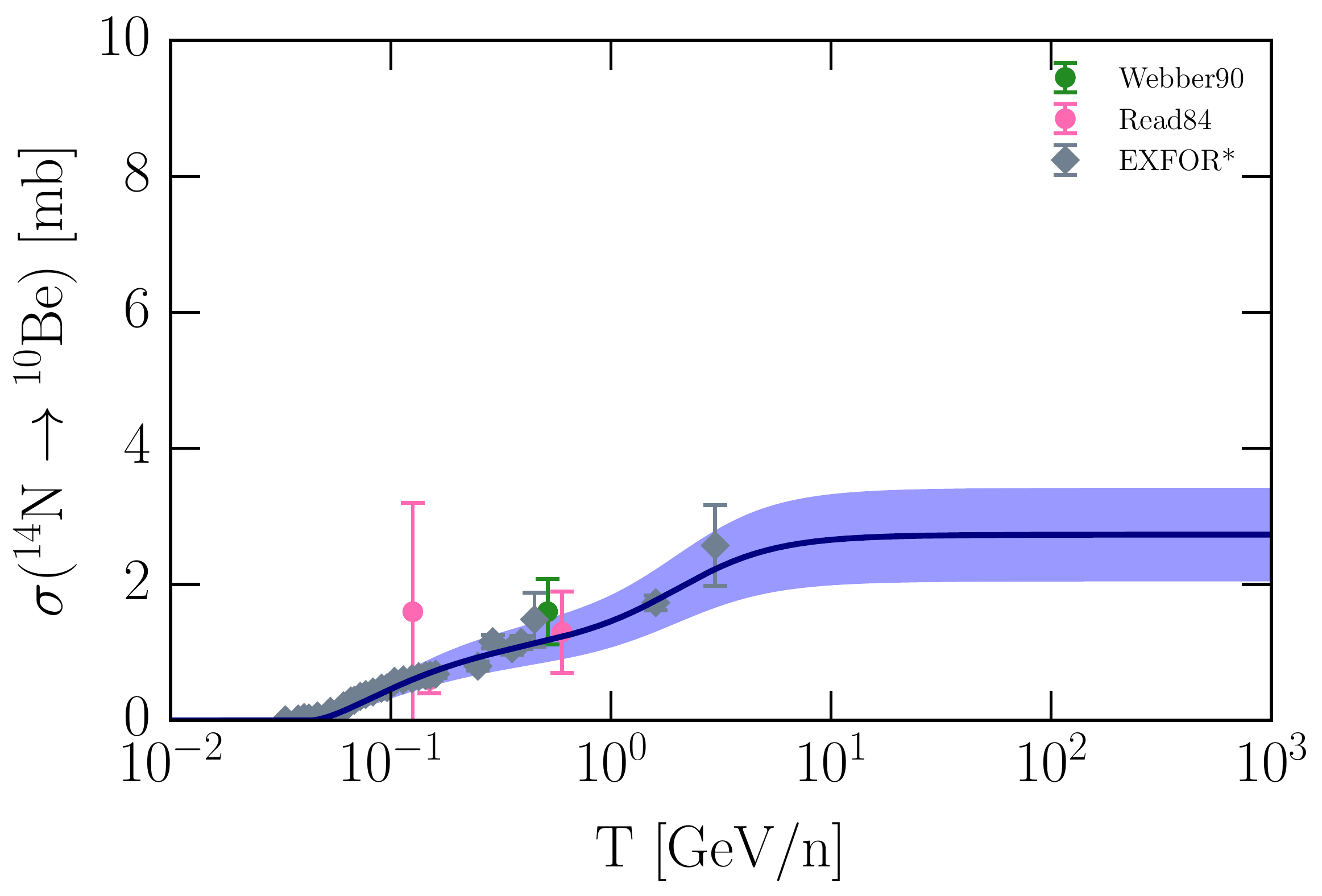} \\
\includegraphics[width=0.31\textheight]{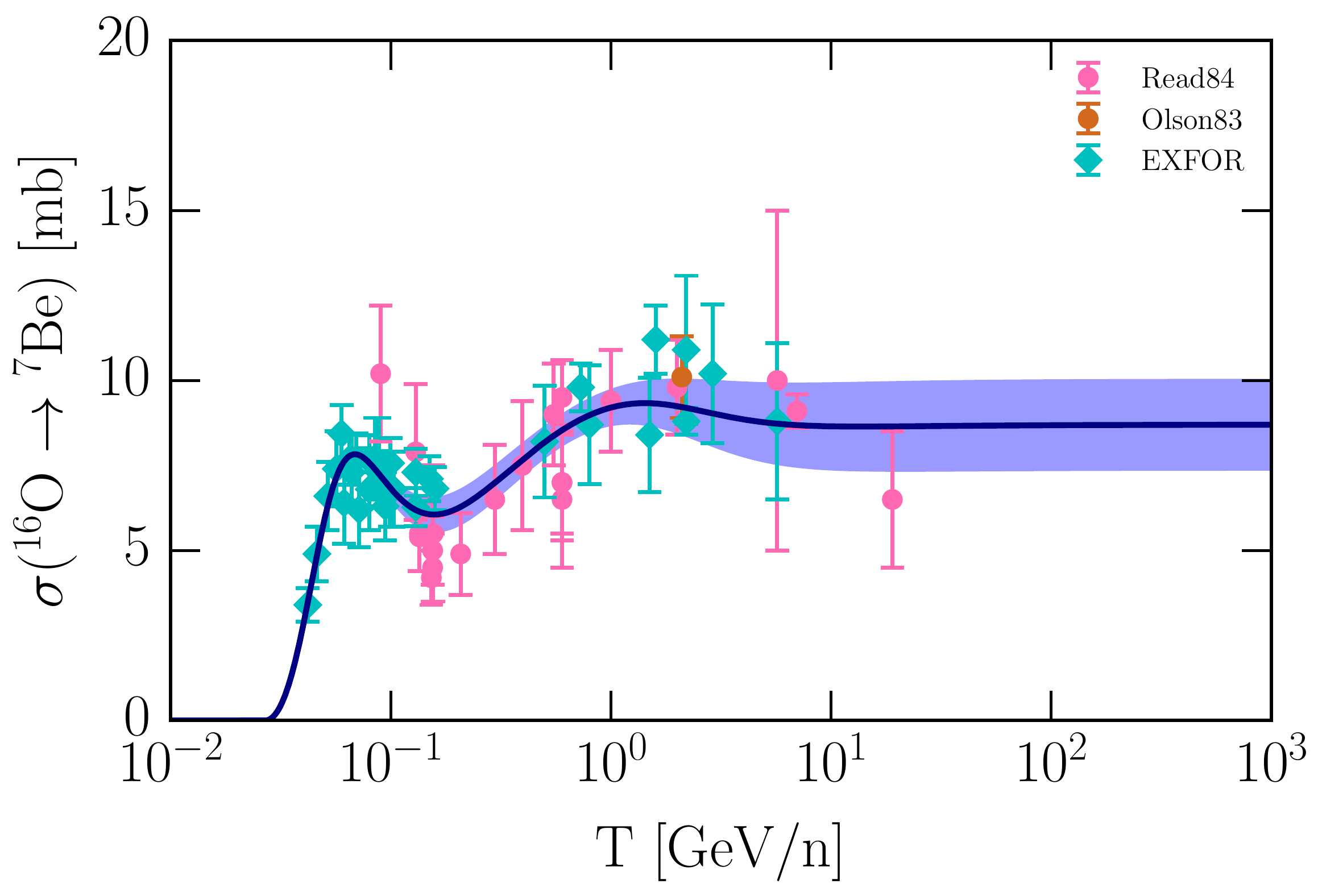}
\includegraphics[width=0.31\textheight]{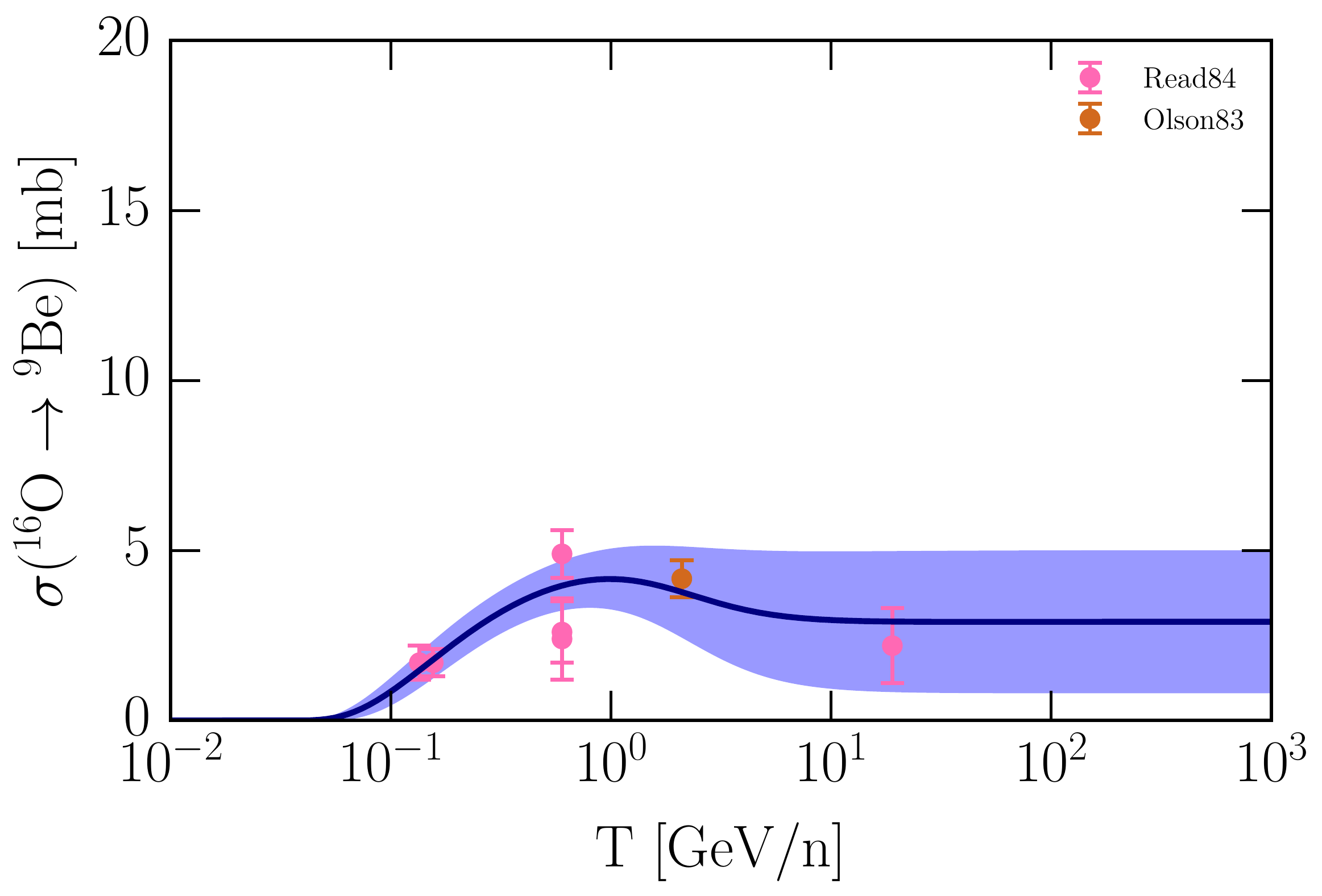} 
\includegraphics[width=0.31\textheight]{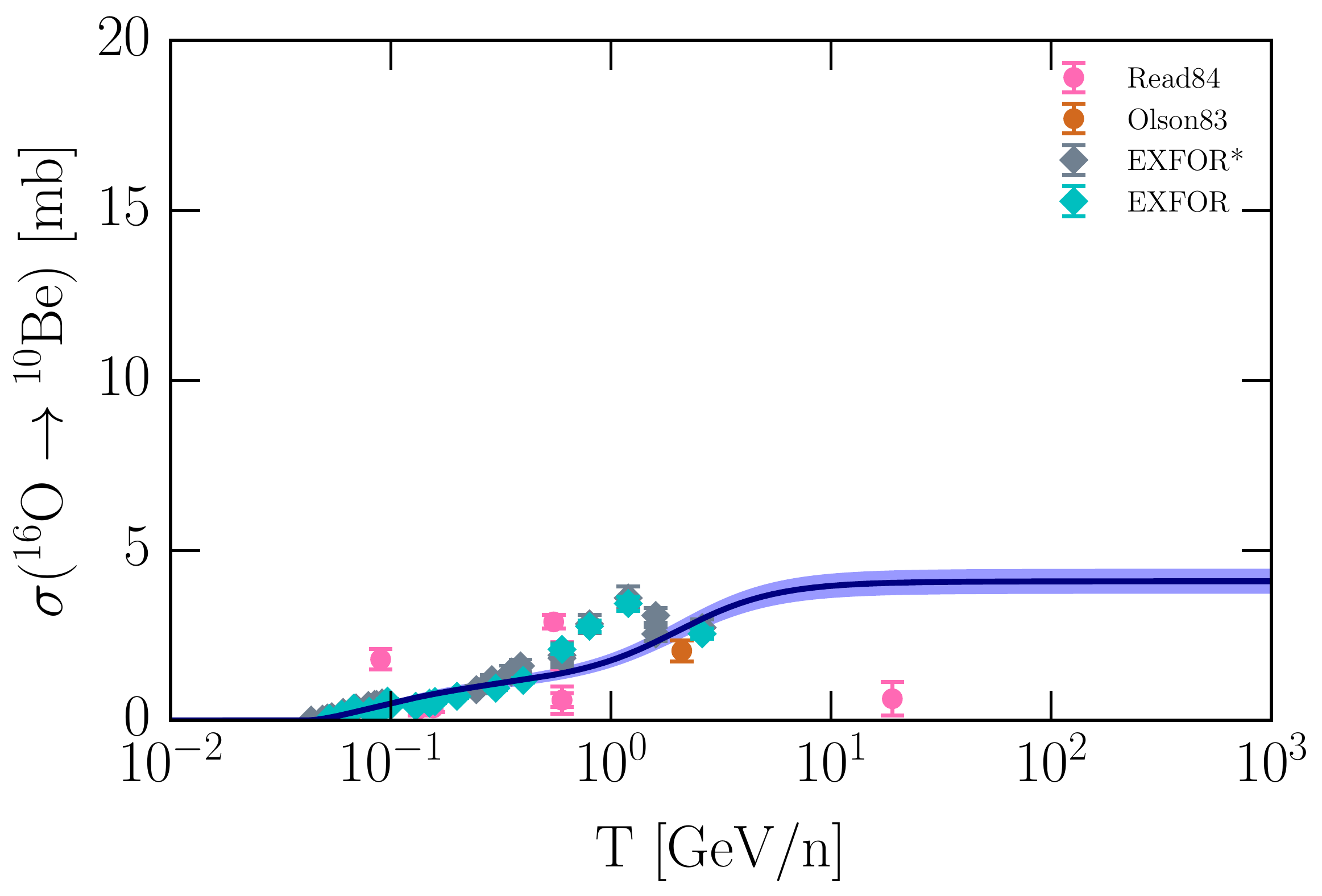} 
\end{center}
\caption{Secondary production cross sections for Berillium isotopes. The lines are labeled as in Fig.~\ref{fig:Lithium}.}
\label{fig:Berillium}
\end{figure}
\end{turnpage}

\begin{turnpage}
\begin{figure}[htp]
\begin{center}
\includegraphics[width=0.31\textheight]{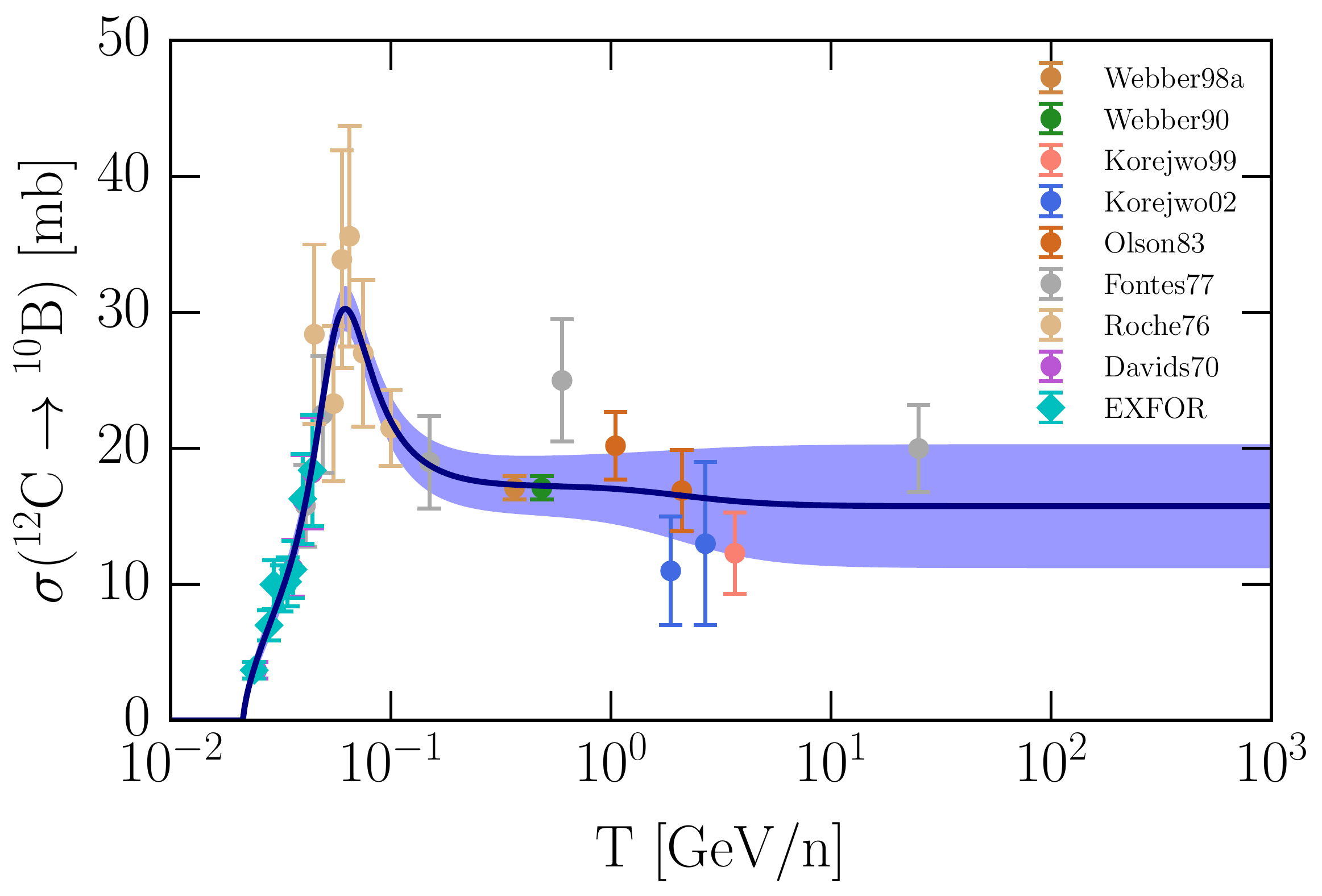} 
\includegraphics[width=0.31\textheight]{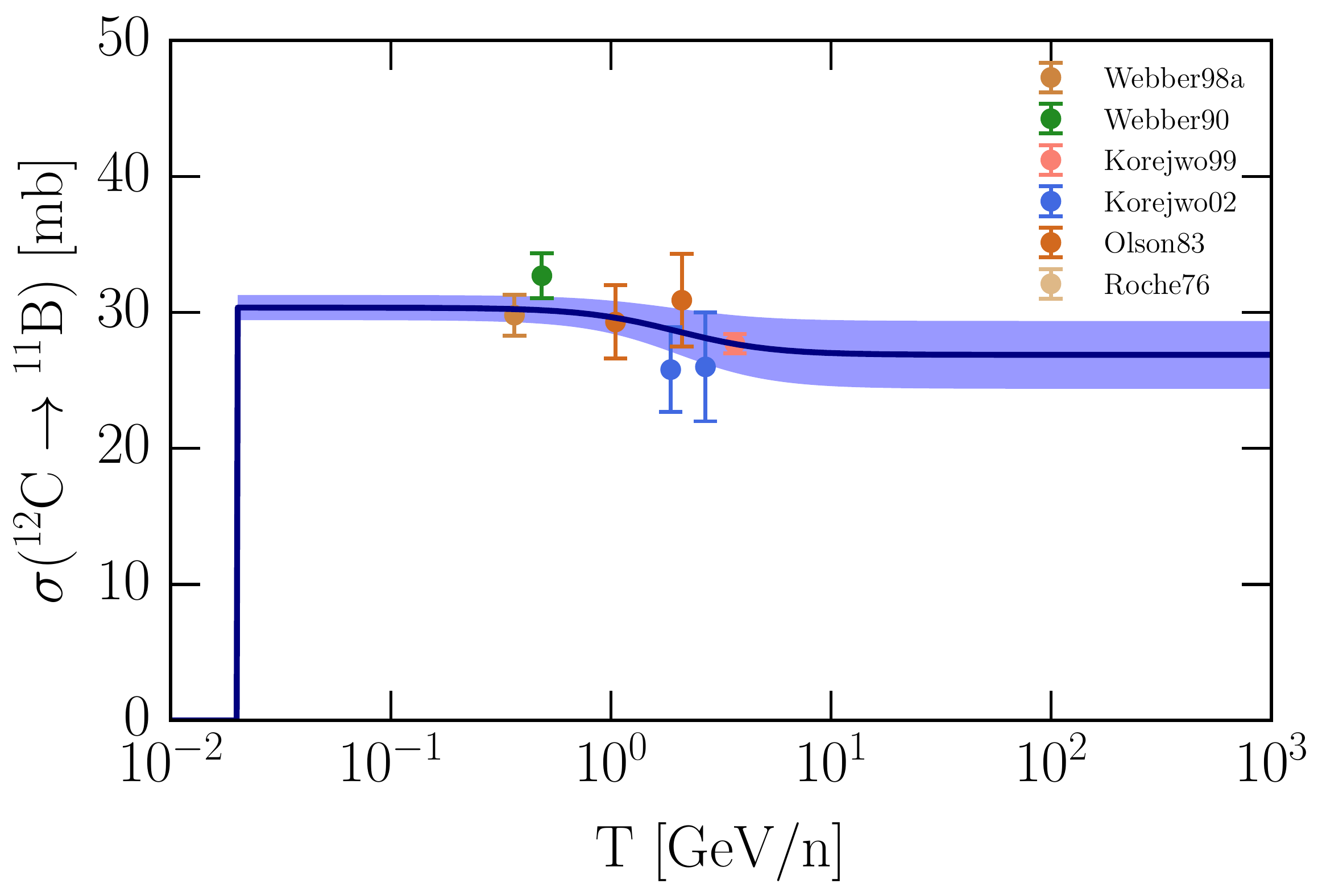}
\includegraphics[width=0.31\textheight]{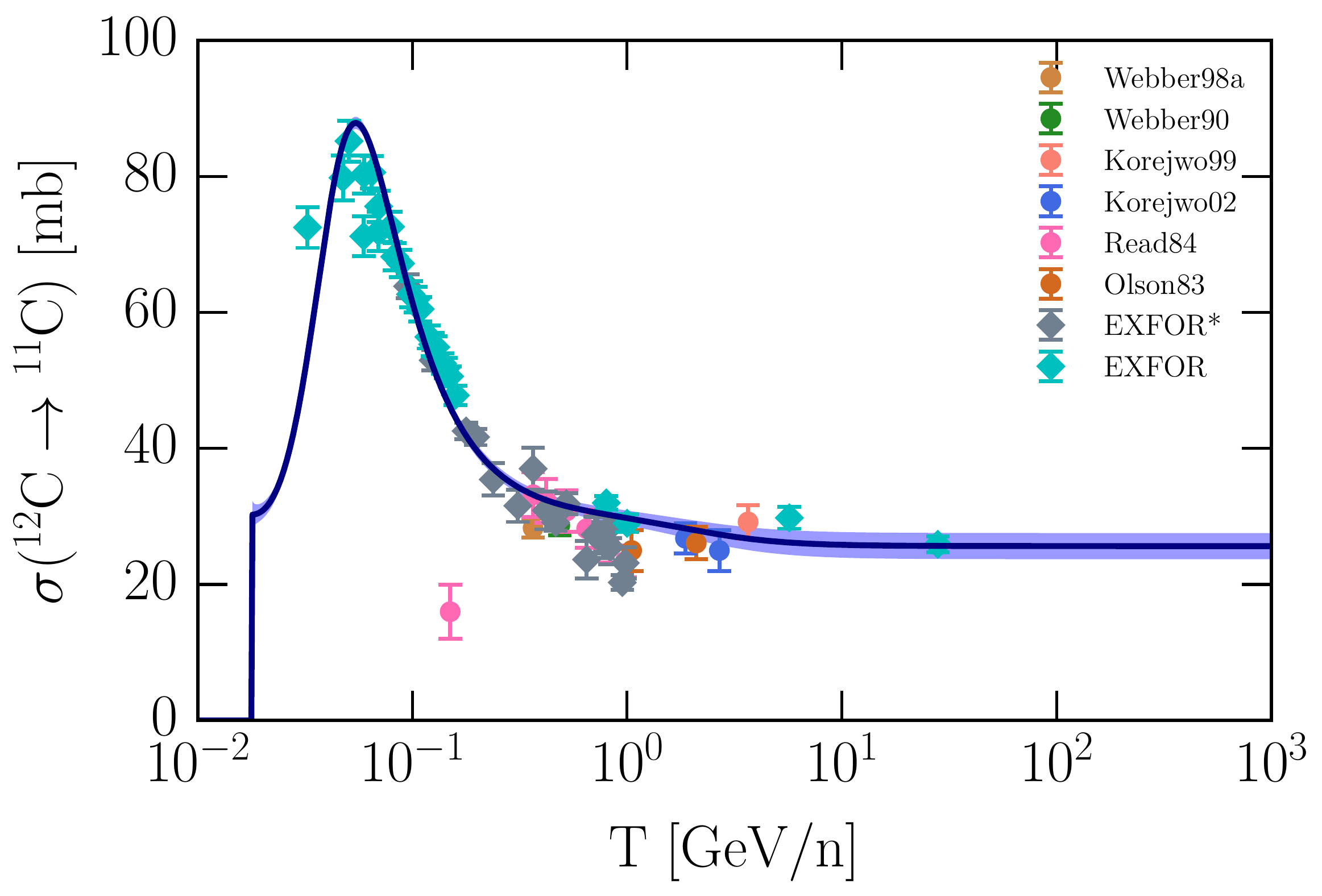} \\
\includegraphics[width=0.31\textheight]{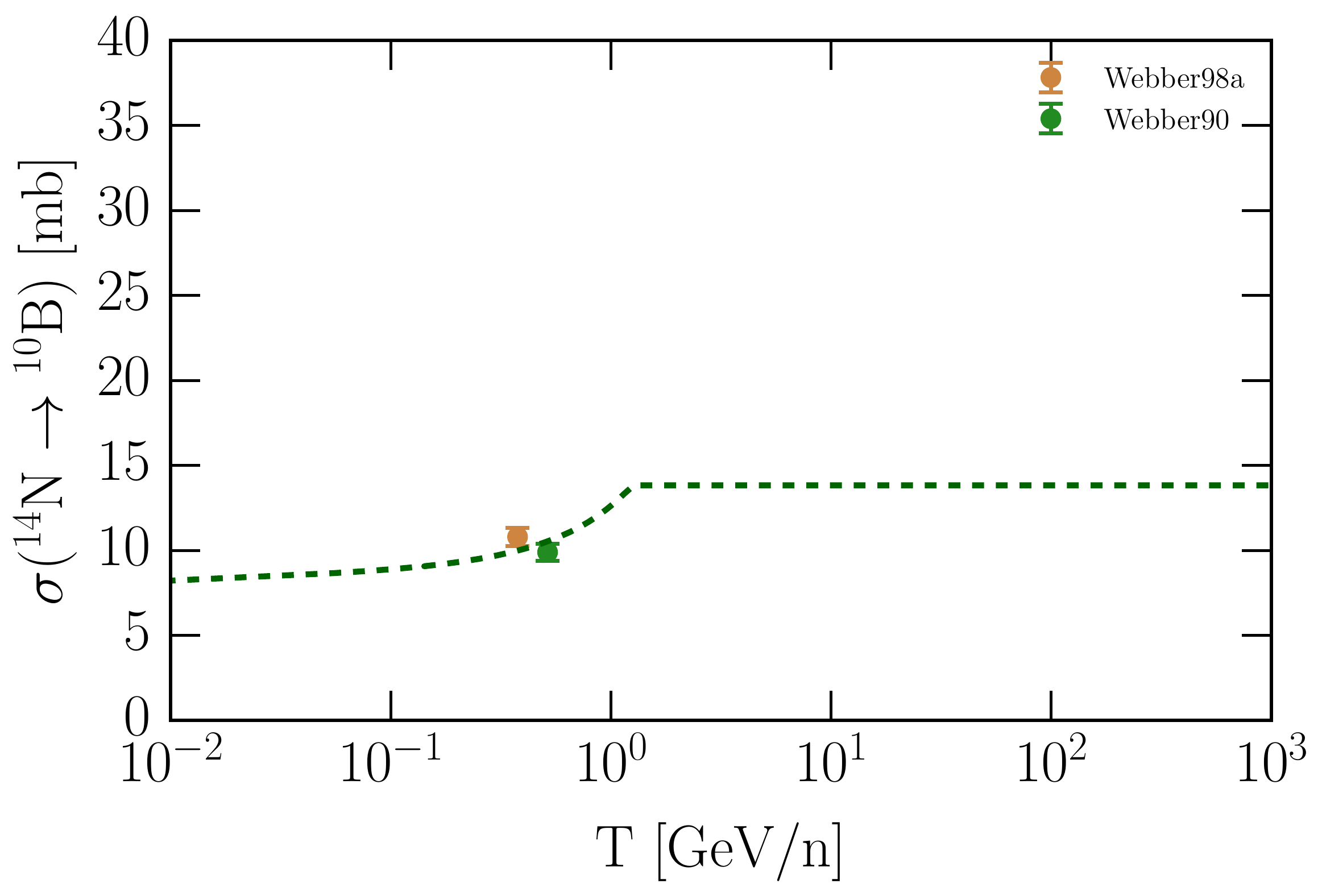} 
\includegraphics[width=0.31\textheight]{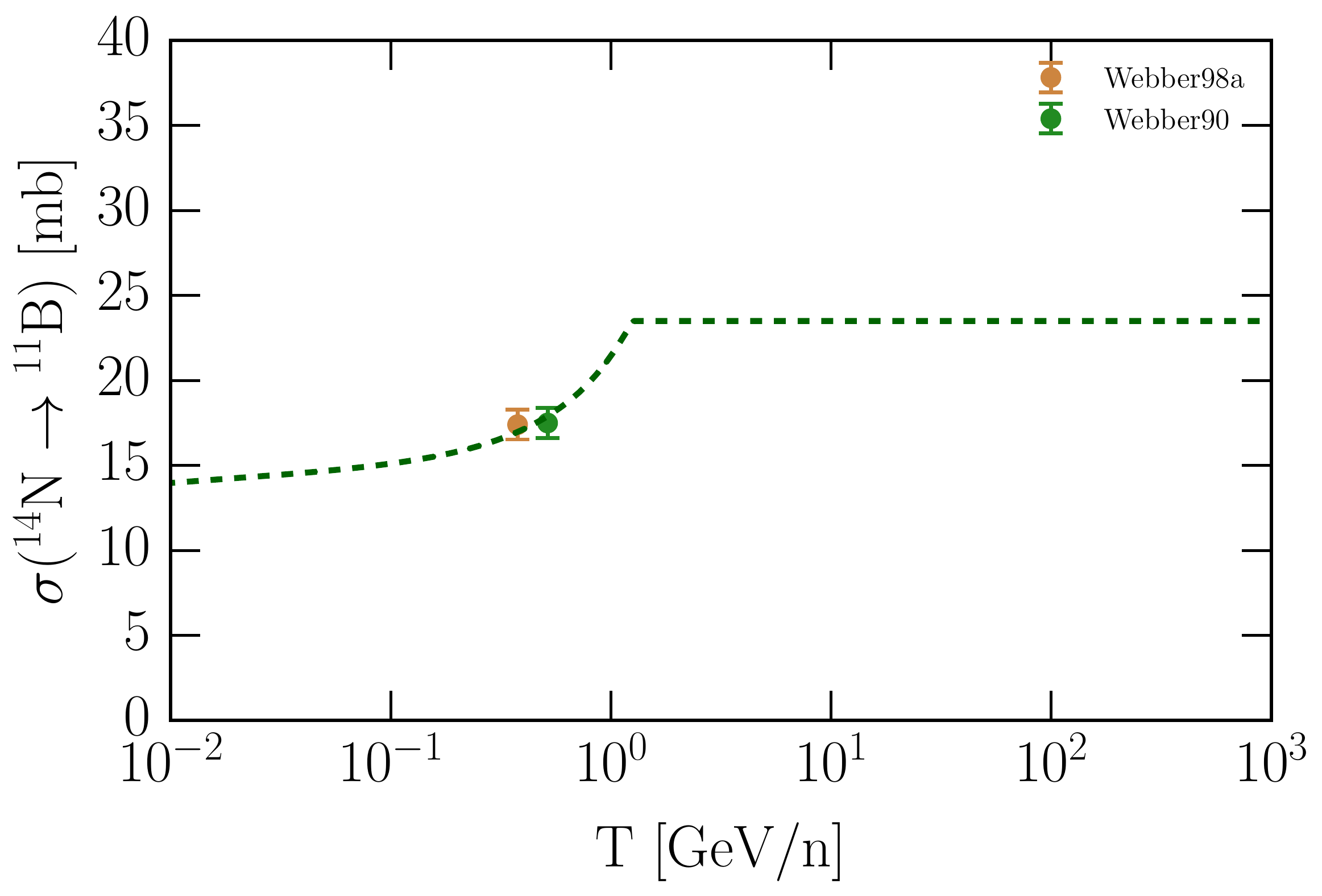} 
\includegraphics[width=0.31\textheight]{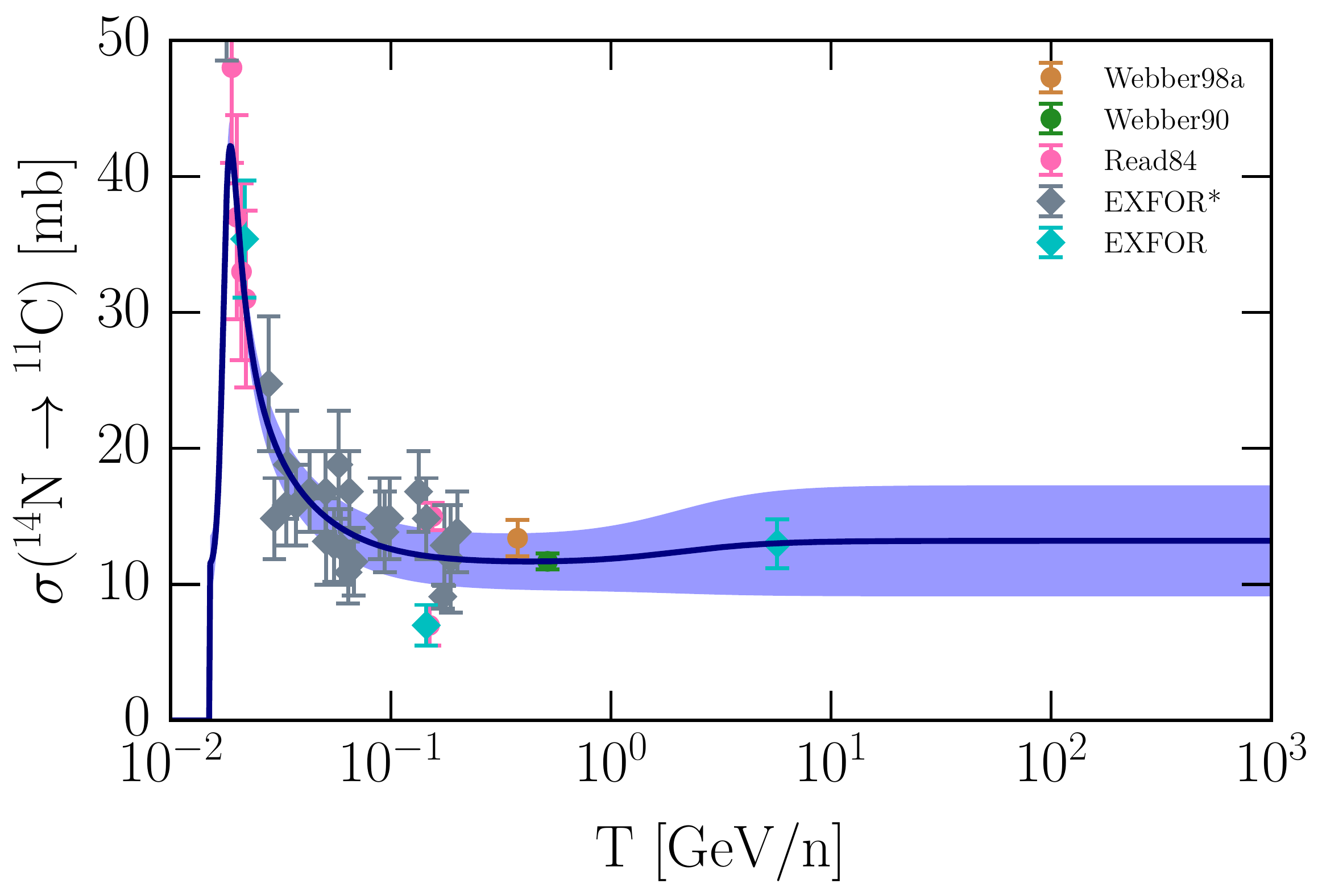} \\
\includegraphics[width=0.31\textheight]{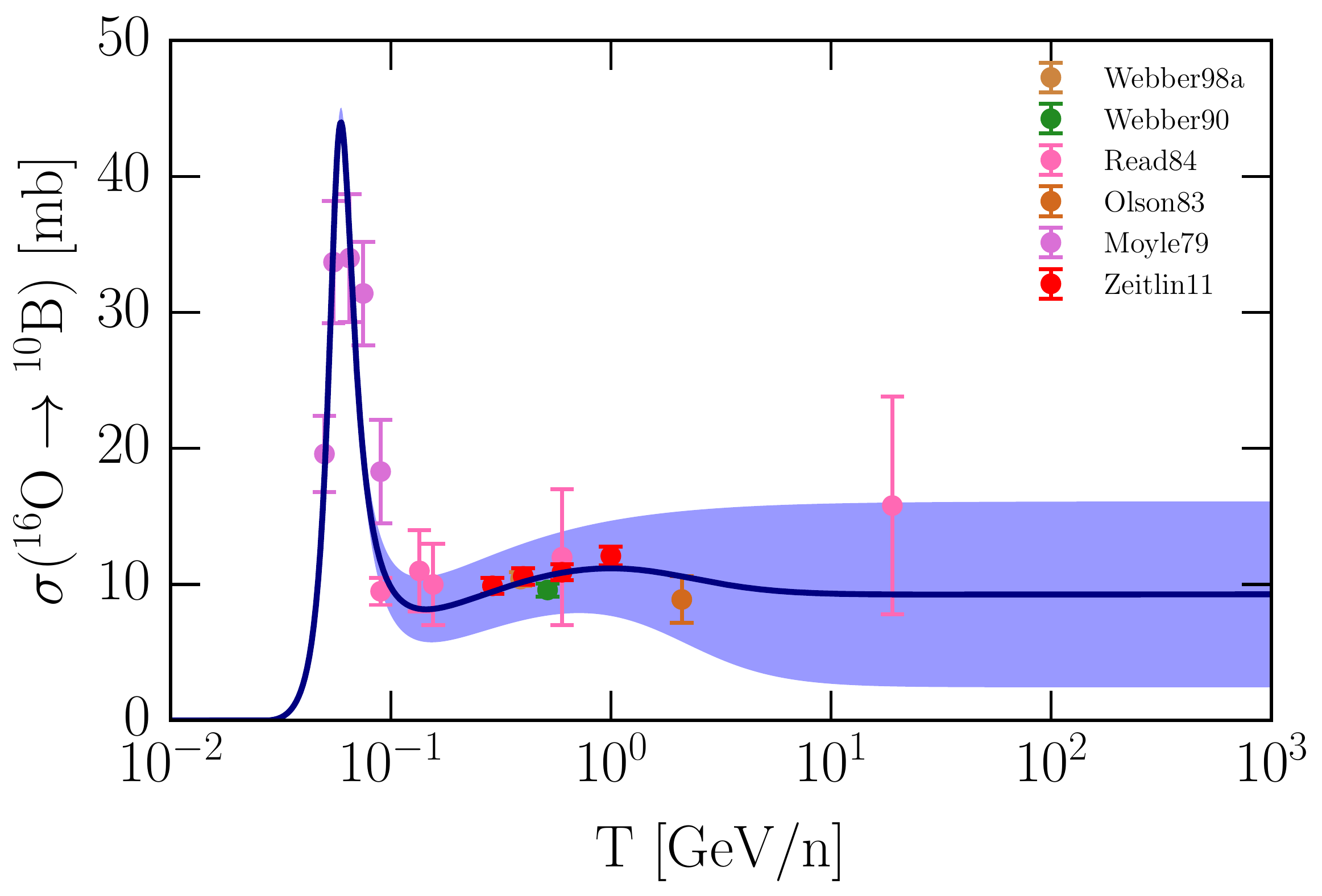} 
\includegraphics[width=0.31\textheight]{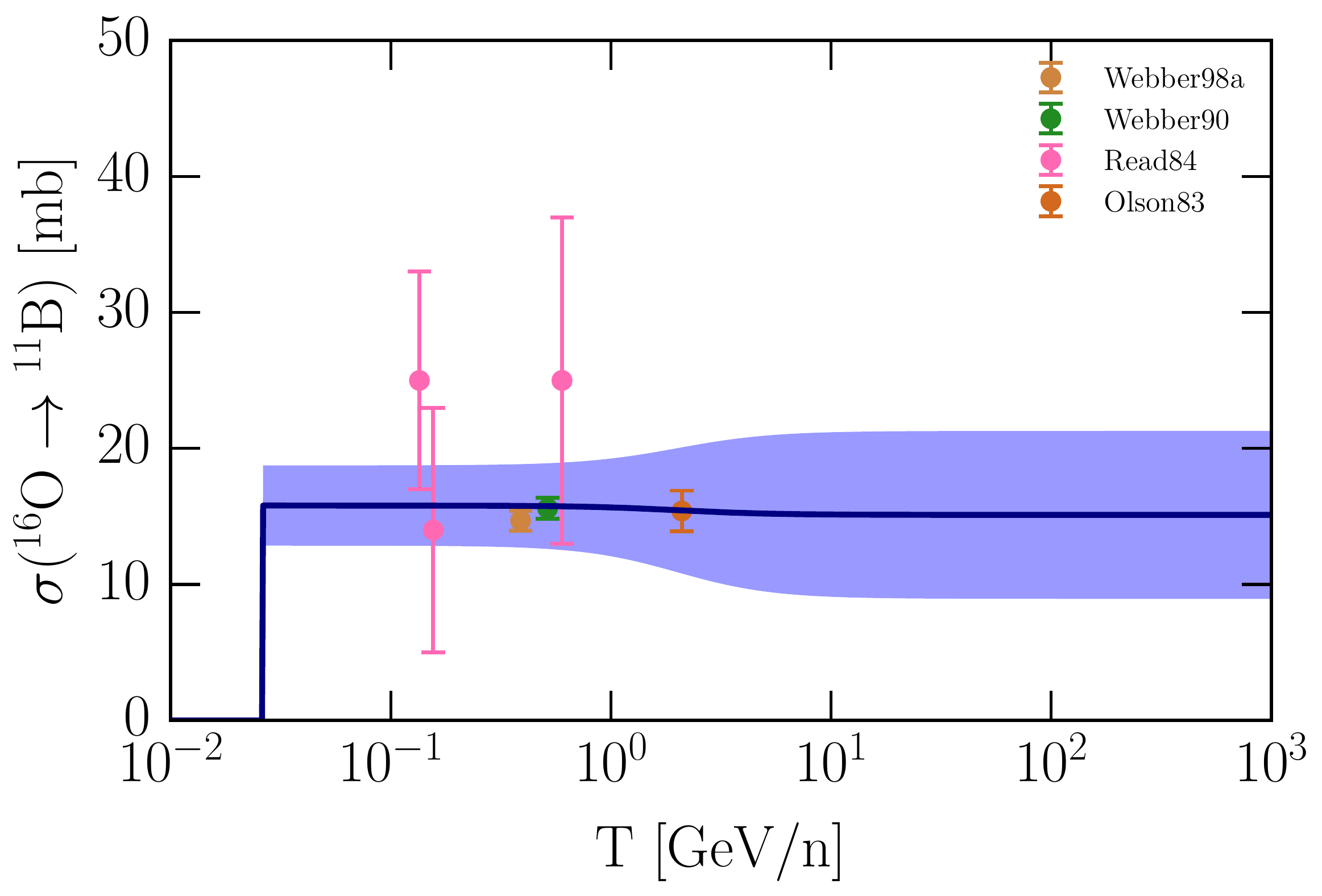} 
\includegraphics[width=0.31\textheight]{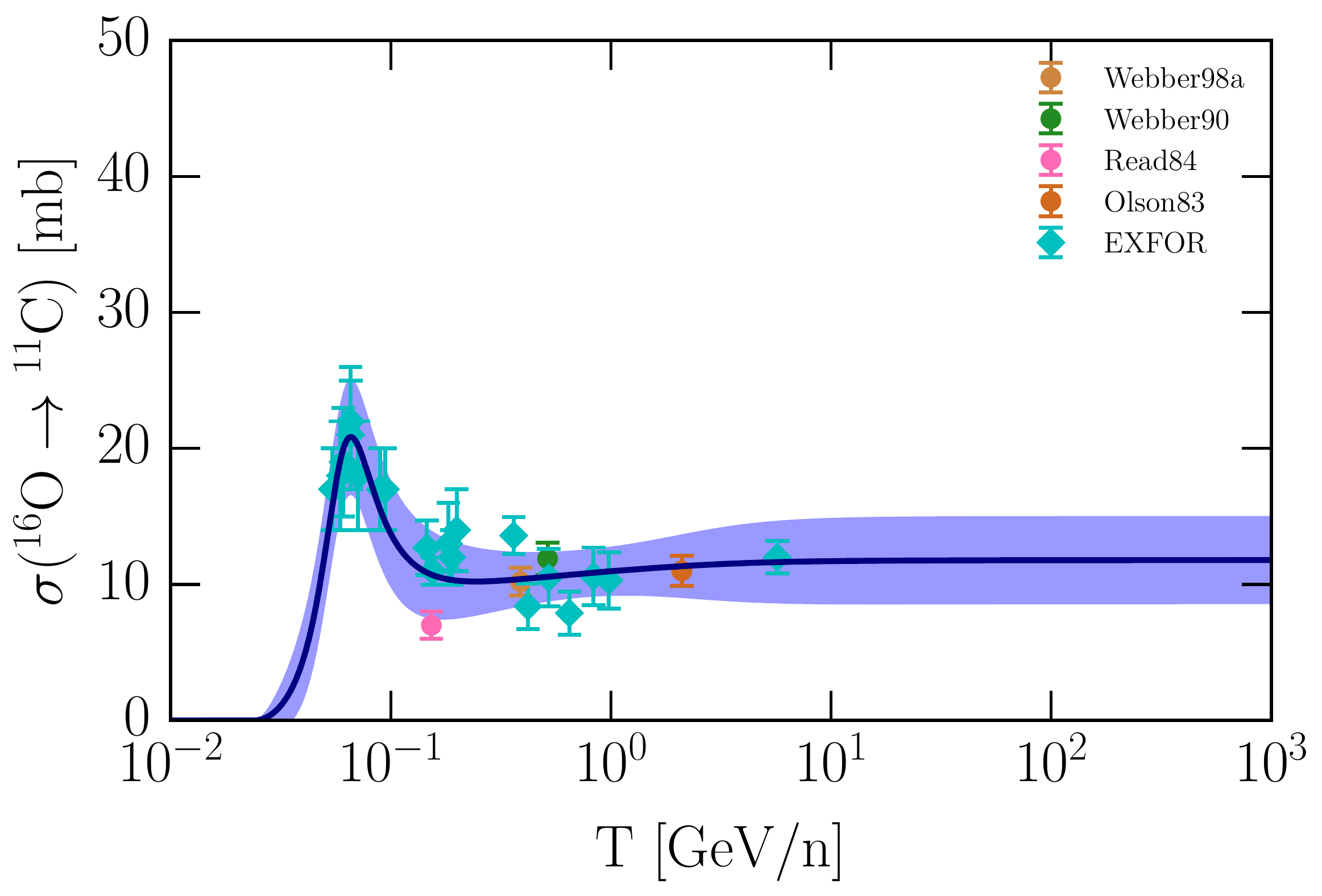} 
\end{center}
\caption{Secondary production cross sections for boron isotopes. The lines are labeled as in Fig.~\ref{fig:Lithium}.}
\label{fig:Boron}
\end{figure}
\end{turnpage}

Figs.~\ref{fig:Lithium}, \ref{fig:Berillium}, and \ref{fig:Boron} show the comparison between the best fit model of fragmentation cross sections and the available data in the energy range from $0.1$ to $10^2$~GeV/n.
We overplot the allowed range for the high-energy value of the cross sections with a shaded region. 
For completeness we also show the channels where, given the sparseness of data, we do not attempt to obtain a fit to our model, but rather decided to use the Webber parametrizations with a suitable renormalization.
  
\newpage

\begin{table*}[htp]
\begin{center}
\begin{tabular}{|cc|cccccccc|}
\hline
\multicolumn{2}{|c|}{Channel} 
& \: $\sigma_{0}$~[mb] \: 
& \: $M$~[MeV] \: 
& \: $\Gamma$~[MeV] \: 
& \: $\sigma_{1}$~[mb] \: 
& \: $\xi$ \:
& \: $\Delta$ \: 
& \: $\left| d\sigma/\sigma \right|$ [\%] \: 
& \: $\chi^2$ \\
\hline 
\multirow{7}{*}{$^{12}$C $\rightarrow$} 
& \color{NavyBlue}{$^{6}$Li} & 20.4 & 40.5 & 8.2 & 12.5(0.9) & 0.56(0.09) & 0.2(0.1) & 20 & 0.85 \\
& \color{NavyBlue}{$^{7}$Li} & & & & 8.2(0.4) & 0(0.1) & 0.7(0.2) & 17 & 2.03 \\
& \color{LimeGreen}{$^{7}$Be} & 66.5 & 31.8 & 43.9 & 10(0.4) & 2.5(0.6) & -0.05(0.05) & 7 & 3.02 \\
& \color{LimeGreen}{$^{9}$Be} & & & & 4.2(0.2) & 0.38(0.05) & 0.7(0.2) & 16 & 0.79 \\
& \color{LimeGreen}{$^{10}$Be} & & & & 3.7(0.2) & 4.2(0.3) & 0.1(0.1) & 17 & 1.69 \\
& \color{RedOrange}{$^{10}$B} & 35.8 & 48.2 & 45.3 & 17.6(0.7) & 0.69(0.06) & -0.1(0.1) & 29 & 0.78 \\
& \color{RedOrange}{$^{11}$B} & & & & 30(1) & 0(0.1) & -0.11(0.05) & 10 & 1.84 \\
& \color{RedOrange}{$^{11}$C} & 44.7 & 19 & 100 & 30.3(0.4) & 0(0.01) & -0.15(0.03) & 8 & 4.42 \\
\hline 
\multirow{2}{*}{$^{14}$N $\rightarrow$} 
& \color{LimeGreen}{$^{7}$Be} & 275.0 & 17.7 & 7.9 & 11.1(0.3) & 5.3(0.6) & 0.10(0.06) & 10 & 3.84 \\
& \color{LimeGreen}{$^{10}$Be} & & & & 1.29(0.07) & 1.9(0.1) & 1.1(0.3) & 25 & 2.00 \\
& \color{RedOrange}{$^{11}$C} & 10$^3$ & 17.5 & 3.7 & 11.5(0.4) & 0(0.006) & 0.1(0.2) & 30 & 27.54 \\
\hline
\multirow{7}{*}{$^{16}$O $\rightarrow$} 
& \color{NavyBlue}{$^{6}$Li} & & & & 10.8(0.9) & 0 & 0.5(0.4) & 60 & 1.71 \\
& \color{NavyBlue}{$^{7}$Li} & & & & 12(3) & 0.5(0.9) & -0.1(0.4) & 82 & 1.64 \\
& \color{LimeGreen}{$^{7}$Be} & 9.6 & 21.7 & 100 & 11.5(0.7) & 6.6(0.8) & -0.24(0.08) & 16 & 0.88 \\
& \color{LimeGreen}{$^{9}$Be} & & & & 5.3(0.9) & 4(1) & -0.4(0.2) & 72 & 1.74 \\
& \color{LimeGreen}{$^{10}$Be} & & & & 1.36(0.05) & 1.94(0.03) & 2.0(0.1) & 10 & 12.21 \\
& \color{RedOrange}{$^{10}$B} & 141.6 & 56.4 & 19.5 & 13(1) & 4(1) & -0.3(0.3) & 74 & 2.39 \\
& \color{RedOrange}{$^{11}$B} & & & & 15.8(0.5) & 0(0.03) & 0(0.2) & 40 & 5.89 \\
& \color{RedOrange}{$^{11}$C} & 37.3 & 52.5 & 44.7 & 11(1) & 2(2) & 0(0.2) & 27 & 1.54 \\  
\hline
\end{tabular}
\end{center}
\caption{Fit parameters entering the fragmentation cross section parametrization~\eqref{eq:fragmentation}.}
\label{tab:fragcross}
\end{table*}

\newpage

\newpage

\newpage

\bibliographystyle{apsrev4-1}
\bibliography{crams}

\end{document}